\documentclass[twocolumn]{aastex63}
\usepackage{graphicx}
\usepackage{float}
\usepackage{hyperref}
\usepackage{natbib}
\usepackage{etoolbox}

\hypersetup{linkcolor=red,citecolor=blue,filecolor=cyan,urlcolor=blue}

\graphicspath{{./}{figures/}}

\newcommand{\sdo}{\textit{SDO}}

\submitjournal{ApJ}

\begin{document}
	
	\title{What Causes Faint Solar Coronal Jets from Emerging Flux Regions in Coronal Holes?}
	\author[0000-0001-7620-3195]{Abigail R. Harden}
	\affiliation{Agnes Scott College, Decatur, GA 30030, USA} 	\email{aharden@agnesscott.edu}

	\author[0000-0001-7620-362X]{Navdeep K. Panesar}
	\affiliation{Lockheed Martin Solar and Astrophysics Laboratory, 3251 Hanover Street, Bldg. 252, Palo Alto, CA 94304, USA}
	\affiliation{Bay Area Environmental Research Institute, NASA Research Park, Moffett Field, CA 94035, USA}
	
		\author[0000-0002-5691-6152]{Ronald L. Moore}
	\affiliation{NASA Marshall Space Flight Center, Huntsville, AL, 35812, USA}
	\affiliation{Center for Space Plasma and Aeronomic Research (CSPAR), UAH, Huntsville, AL 35805, USA}
	
	\author[0000-0003-1281-897X]{Alphonse C. Sterling}
	\affiliation{NASA Marshall Space Flight Center, Huntsville, AL, 35812, USA}
	
		\author{Mitzi L. Adams}
	\affiliation{NASA Marshall Space Flight Center, Huntsville, AL, 35812, USA}

	\begin{abstract}

		   Using EUV images and line-of-sight magnetograms from \textit{Solar Dynamics Observatory}, we examine eight emerging bipolar magnetic regions (BMRs) in central-disk coronal holes for whether the emerging magnetic arch made any noticeable coronal jets directly, via reconnection with ambient open field as modeled by \cite{yokoyama95}.  During emergence, each BMR produced no obvious EUV coronal jet of normal brightness, but each produced one or more faint EUV coronal jets that are discernible in AIA 193 \AA\ images.  The spires of these jets are much fainter and usually narrower than for typical EUV jets that have been observed to be produced by minifilament eruptions in quiet regions and coronal holes.  For each of 26 faint jets from the eight emerging BMRs, we examine whether the faint spire was evidently made a la \cite{yokoyama95}.  We find: (1) 16 of these faint spires evidently originate from sites of converging opposite-polarity magnetic flux and show base brightenings like those in minifilament-eruption-driven coronal jets, (2) the 10 other faint spires \textit{maybe} were made by a burst of the external-magnetic-arcade-building reconnection of the emerging magnetic arch with the ambient open field, reconnection directly driven by the arch’s emergence, but (3) \textit{none} were unambiguously made by such emergence-driven reconnection.  Thus, for these eight emerging BMRs, the observations indicate that emergence-driven external reconnection of the emerging magnetic arch with ambient open field \textit{at most} produces a jet spire that is much fainter than in previously-reported, much more obvious coronal jets driven by minifilament eruptions.
		
		\end{abstract}
	

	\section{Introduction} \label{sec:intro}

	A solar coronal jet is a sudden plasma outflow from a compact base of closed magnetic field that stands on the Sun’s surface and is hemmed in by surrounding unipolar magnetic field that reaches much higher up into the corona.  The base is typically less than 30,000 km across, smaller than a supergranule.  An episode of reconnection of the closed base field with the ambient far-reaching field drives a magnetically-channeled transient narrow ejection of magnetized plasma – usually no wider than the base – up along the ambient field into the corona \citep{Shibata_1992,Moore_2010,Raouafi_2016,innes16}.  In this paper, we call the narrow column of ejected plasma the jet’s “spire.”  We use the term “jet” interchangeably with “spire” when it is clear from the context that we are referring to only the spire, the narrow ejected outflow.  Otherwise, the term “jet” refers to the entire event, encompassing both the spire and what happens at the base to drive the spire.  Because plasma in the spire and in the base of a coronal jet is usually heated to transition-region and coronal temperatures (temperatures from $\sim$ 10$^{5}$ K to a few times 10$^{6}$K), coronal jets are typically seen in extreme ultraviolet (EUV) images \citep{Nistico_2009,Panesar_2016a,Sterling_2016}, coronal X-ray images \citep{Shimojo&H_1996,Cirtain_2006,Savcehva_2007}, or both \citep{Moore_2010,Moore_2013,Sterling_2015}.

		Coronal jets can occur only once or repeatedly at the same place, up to more  than ten in  less than  a day  \citep[e.g.][]{liu16,Panesar_2017}.  The base always includes an island of magnetic flux of one polarity embedded in surrounding predominantly unipolar field of the opposite polarity.  Jets originate at such places in coronal holes, in quiet regions, and at the edges of sunspot active regions \citep{Savcehva_2007,schmieder13,Panesar_2016a,Sterling_2016}.  In each case, the jet base is a fan of magnetic loops that connect the flux island to the surrounding flux of opposite polarity.  The polarity of the island’s flux is the “minority” polarity, and the polarity of the surrounding flux is the “majority” polarity.   
	
	Coronal jets usually have in the base a compact brightening that is much smaller than the whole base and sits near an edge of the base \citep{Shibata_1992}.  Following \cite{Sterling_2015}, we call the base-edge compact bright feature the “jet bright point” or its acronym, the JBP.  Throughout the spire’s lifetime of typically $\sim$ 10 minutes, the spire usually sits off the JBP, and usually drifts laterally away from the JBP \citep{savcheva09,Sterling_2015}.
	
	It is generally thought that the spire is the upward-outflow product of external reconnection of the jet-base closed field, that is, reconnection of the outside of jet-base closed field with oppositely-directed ambient far-reaching field at a current-sheet interface at which these two fields push against each other.  What drives the formation of the interface and the consequent spire-making reconnection remains a debated question.  That is the underlying question motivating the work reported here.

	    There are presently two main alternative ideas – one older, the other more recent – for the driver of the closed-field/far-reaching-field reconnection that produces the spire.  The older idea is based on the observationally-suggested supposition that, at the time of a coronal-jet eruption, much of the jet-base closed field is the magnetic arch of a bipolar magnetic region (BMR) that is emerging through the photosphere from below.  On this basis, with support from numerical simulations of coronal-jet production by such magnetic arch emergence, the older idea is that it is the ongoing emergence of the BMR’s magnetic arch that pushes the outside of the arch against oppositely-directed ambient far-reaching field, thereby driving the arch’s external reconnection that makes the spire \citep[e.g.][]{Shibata_1992,YokoyamaShibata_1996,Moreno-Insertis_2008}.  The more recent idea is based on observations indicating that, before a coronal jet occurs, the core of a lobe of the jet-base closed field
	is sheared and twisted into a flux rope along the lobe’s polarity inversion line (PIL).  A miniature filament, called a “minifilament,” usually resides in and reveals the sheared and twisted core field.  In the manner of full-size filaments seen in H$\alpha$ images \citep{martin86, martin98}, the minifilament is PIL-tracing magnetically suspended cool plasma seen as a dark feature in arcsecond-resolution EUV images \citep[e.g.][]{adams14,Sterling_2017,Panesar_2016b,Panesar_2018}.  On this basis, and from observed erupting minifilaments in coronal  jets, the more recent idea is that it is the eruption of the minifilament flux rope and its enveloping lobe of the closed-field jet base that pushes the outside of the erupting lobe against oppositely-directed ambient far-reaching field, thereby driving the lobe’s external reconnection that makes the spire \citep[e.g.][]{Sterling_2015,Panesar_2016b,Panesar_2017,Sterling_2017,Panesar_2018,moore18,panesar20,McGlasson_2019}. By each being based on at least ten observed jets, the papers cited in the previous sentence established the viability of the more recent idea. Earlier papers reporting coronal jets involving minifilament eruptions were each limited to a single jet \cite[e.g.][] {Hong_2011,shen12,Huang_2012,adams14}.

	The two alternative ideas for coronal-jet production also starkly differ in how the JBP is made.  In the flux-emergence idea, the JBP is the hot closed-loop downward product of the emerging arch’s external reconnection, the upward product of which is the spire.  It is supposed that little external reconnection of the emerging arch with the ambient far-reaching field has occurred prior to the burst of external reconnection that makes both the spire and the new hot closed loops that are taken to be the JBP.  If that is true, then the new hot closed loops made by the spire-making external reconnection are in agreement with observed JBPs in being located at an edge of the base and much smaller than the whole base.  In the flux-emergence picture, most of the base is the emerging arch.  Thus, in the flux-emergence idea, before, during, and after a jet eruption, the form of the closed-field base is that of a very lopsided anemone that splays out from the patch of emerged minority-polarity flux.  The anemone is made of two lobes, one much smaller than the other.  The big lobe is the emerging magnetic arch.  During the emergence prior to the jet, a small amount of external reconnection of the emerging arch with the ambient field makes the little lobe.  During the jet, the little lobe is made bigger by the hot new loops (supposedly the JBP) added to its outside by the spire-making burst of external reconnection of the emerging arch, but remains much smaller than the emerging arch.

		In the minifilament-eruption idea, before and after a jet, the jet base is closed field having the same topology as the jet base in the flux-emergence idea.  The base field is again a lopsided anemone made of two back-to-back lobes, a larger one and a smaller one, each stemming from the minority-polarity flux patch inside the anemone.  The minifilament flux rope is in the core of the smaller lobe.  It is the eruption of the minifilament-carrying smaller lobe that drives the production of the jet.  The erupting lobe drives external reconnection of itself with encountered oppositely-directed far-reaching field.  That reconnection simultaneously produces the spire and adds new hot loops to the outside of the anemone’s other lobe, the larger one.  Because the larger lobe is most of the closed-field base, the new hot loops added to the larger lobe by the external reconnection of the erupting lobe cannot make a compact EUV/X-ray JBP at an edge of the base.  Instead, they make widespread EUV/X-ray brightening in the base.  In the minifilament-eruption idea, the JBP is a small hot-loop arcade made by internal reconnection of the minifilament-flux-rope-cored erupting lobe, in the manner of the flare arcade that is made by reconnection of the legs of a large filament-flux-rope-cored erupting magnetic arcade that becomes a coronal mass ejection (CME).  That is, in the minifilament-eruption idea for coronal-jet production, the minifilament eruption is a small-scale version of the filament eruptions that make CMEs, and the JBP is a small-scale version of the flare arcade produced in CME-making filament eruptions \cite[e.g.][]{sterling18}.

     \cite{Panesar_2016b,Panesar_2017,Panesar_2018,McGlasson_2019}  combined, found that at least a large majority ($>$80\%) of $\sim$ 90 obvious coronal jets were clearly driven by a minifilament eruption in the manner described above. They found this from close inspection of each jet and its preamble in arcsecond-resolution coronal EUV movies and co-aligned arcsecond-resolution magnetograms of the jet-base magnetic flux and its pre-jet evolution.  They selected each jet for study only because it was eye-catchingly obvious in three-minute-cadence movies of full-disk arcsecond-resolution coronal EUV images and happened in a quiet region or coronal hole on the central disk.  Therefore, the set of $\sim$90 jets that they studied is a random sample of obvious coronal EUV jets seated in quiet regions and coronal holes on the central disk.

For observed minifilament-eruption-driven jets, magnetograms show that the jet base sits on a flux patch of one polarity (the minority polarity) in a predominantly-unipolar larger region of flux of the opposite polarity (the majority polarity). Coronal images co-aligned with the magnetograms show that the jet-base closed field is a magnetic anemone (the afore-mentioned full-circle fan of magnetic loops) that encircles the minority-polarity flux and arches over the PIL encircling that flux  \citep{Panesar_2016b}.  The anemone connects the minority-polarity flux to the surrounding majority-polarity flux.  Tracking the minority-polarity flux patch backward in time in magnetograms shows that the minority-polarity flux patch originates hours to days earlier.  It originates either as an emerging BMR’s minority-polarity flux or by coalescence of minority-polarity flux that was near or below the sensitivity threshold of the magnetograms \citep{Panesar_2017}.  Regardless of the origin of the jet-base anemone’s minority-polarity flux, as photospheric flow gradually carries that flux closer to the anemone’s majority-polarity flux on one side of the anemone, the anemone’s lobe rooted in the converging flux gradually shrinks, and the minifilament and the sheared field that holds it start building up along the PIL between the converging opposite-polarity flux.

Presumably, the minifilament-traced sheared field and flux rope are built along the PIL as follows, in the way that \cite{moore84} conjectured and \cite{balle89} depicted and numerically simulated.  The field in the shrinking lobe of the anemone has some initial shear.  The photospheric-flow-driven convergence of the lobe’s flux toward the shrinking lobe’s PIL amplifies the field’s shear, concentrates the shear along the PIL, and builds the flux rope by driving reconnection of the sheared field low above the PIL.  As the reconnection increases the shear and makes the flux rope, the reconnection also makes short PIL-crossing loops.  Submergence of the short loops constitutes flux cancelation at the PIL.  Evidently, continuation of such photospheric-flow-driven reconnection and flux cancelation at the PIL under the minifilament eventually triggers the jet-driving eruption of the minifilament flux rope and its enveloping lobe of the base anemone \citep{moore92,Young&Muglach_2014,Panesar_2017}.

	
	\begin{longrotatetable}
		\begin{deluxetable*}{c c c c c c c c c c c c}
			\tablewidth{0pt}
			\tabletypesize{\footnotesize}
			\renewcommand{\arraystretch}{1.6}
			\setlength{\tabcolsep}{3.0pt} 
			\tablenum{1}
			\tablecaption{The Emerging Bipolar Magnetic Regions and their Faint Jets \label{tb: params}}
			\tablehead{
				\colhead{Emerging} & 
				\colhead{Start Time\tablenotemark{\scriptsize a}} & 
				\colhead{End Time\tablenotemark{\scriptsize b}} & 
				\colhead{Location\tablenotemark{\scriptsize c}} &
				\colhead{No. of\tablenotemark{\scriptsize d}} & 
				\colhead{Jet} & 
				\colhead{Time\tablenotemark{\scriptsize e}} &
				\colhead{Spire\tablenotemark{\scriptsize f}} & 
				\colhead{Spire\tablenotemark{\scriptsize g}} &
				\colhead{Base\tablenotemark{\scriptsize h}} &
				\colhead{Base\tablenotemark{\scriptsize i}} & 
				\colhead{Spire Dr.\tablenotemark{\scriptsize j}} 
				\\\colhead{BMR}&
				\colhead{(UT)} & 
				\colhead{(UT)}&
				\colhead{x,y}&
				\colhead{Faint}&
				\colhead{No.}&
				\colhead{(UT)}&
				\colhead{Duration}&
				\colhead{Speed} &
				\colhead{bri.?} &
				\colhead{bri. at} &
				\colhead{by BMR} 
				\\\colhead{Region No.}&
				\colhead{}&
				\colhead{}&
				\colhead{(arcsec)}&
				\colhead{Jets}&
				\colhead{}&
				\colhead{}&
				\colhead{(min.)}&
				\colhead{(km s $^{-1}$)}&
				\colhead{}&
				\colhead{can. site?}&
				\colhead{Emerg.?}
			}
			\startdata
			E1 & 21:00 2012 June 30 & 05:00 2012 July 1 & 126,282 & 1 & J1 & 21:30 & 2 $\pm$ 1 & 90 $\pm$ 30 & Y & N & A \\ 
			E2 & 04:00 2012 July 1  & 15:00 2012 July 1 & 244,7 & 3 & J2 & 09:33 & 10 $\pm$ 1 & 101 $\pm$ 6 & Y & Y & A \\ 
			&  &  &  &  & J3 & 09:48 & 15 $\pm$ 5 & 90 $\pm$ 5 & Y & N & A\\ 
			&  &  &  &  & J4 & 10:52 & 13 $\pm$ 4 & 90  $\pm$ 10 & Y & N & A\\
			E3 & 15:00 July 1 2012 & 08:00 July 2 2012 & 390,-282 & 2 & J5\tablenotemark{\scriptsize k} & 18:17 & 9 $\pm$ 2 & -\tablenotemark{\scriptsize l} & Y & Y & N\\ 
			&  &  &  &  & J6\tablenotemark{\scriptsize k} & 18:41 & 4 $\pm$ 1 & -\tablenotemark{\scriptsize l} & Y & Y & N\\
			E4 & 20:00 2015 January 9 & 18:00 2015 January 11 & 13,59 & 1 & J7 & 12:45 & 7 $\pm$ 5 & 180 $\pm$ 65 \tablenotemark{\scriptsize m} & Y & Y & N\\ 
			E5 & 07:00 2015 January 10 & 15:30 2015 January 10 & 116,-10 & 10 & J8\tablenotemark{\scriptsize n} & 10:19 & 9 $\pm$ 2 & 235 $\pm$ 70& Y & Y & N\\ 
			&  &  &  &  & J9\tablenotemark{\scriptsize n} & 10:31 & 6 $\pm$ 3 & 200 $\pm$ 10 & Y & Y & N\\
			&  &  &  &  & J10\tablenotemark{\scriptsize o} & 10:40 & 2 $\pm$ 1 & 120 $\pm$ 30 & Y & Y & N\\
			&  &  &  &  & J11\tablenotemark{\scriptsize o} & 10:42& 9 $\pm$ 2 & 160 $\pm$ 10 & Y & Y & N\\
			&  &  &  &  & J12\tablenotemark{\scriptsize p} & 10:58 & 11 $\pm$ 3 & 115 $\pm$ 10 & Y & Y & N\\
			&  &  &  &  & J13\tablenotemark{\scriptsize p} & 11:10 & 8 $\pm$ 3 & 100 $\pm$ 10 & Y & Y & N\\
			&  &  &  &  & J14 & 11:23 & 13 $\pm$ 3 & 110 $\pm$ 10 & Y & Y & N\\
			&  &  &  &  & J15\tablenotemark{\scriptsize q} & 12:09 & 8 $\pm$ 2 & 130 $\pm$ 10 & Y & Y & N\\
			&  &  &  &  & J16\tablenotemark{\scriptsize q} & 12:20 & 8 $\pm$ 1 & 120 $\pm$ 10 & Y & Y & N\\
			&  &  &  &  & J17 & 17:24 & 10 $\pm$ 4 & 35 $\pm$ 10\tablenotemark{\scriptsize m} & Y & Y &  N\\
			E6 & 06:00 2017 January 31 & 06:00 2017 February 1 & -53,346 & 3 & J18 & 08:16 & 4 $\pm$ 2 &  85 $\pm$ 10& Y & Y & N\\ 
			&  &  &  &  & J19 & 08:26 & 10 $\pm$ 3 & 145 $\pm$ 5 & Y & Y &N\\
			&  &  &  &  & J20 & 09:04 & 7 $\pm$ 1 & 135 $\pm$ 10 & Y & Y & N\\
			E7 & 03:30 2011 February 27 & 04:10 2011 February 27 & -337,391 & 1 & J21 & 04:31 & 11 $\pm$ 5 & 80 $\pm$ 20 & Y & N & A\\ 
			E8 & 21:40 2012 May 12 & 01:30 2012 May 13 & 60,573 & 5 & J22 & 22:13 & 4 $\pm$ 1 & -\tablenotemark{\scriptsize l} & Y & N & A\\ 
			&  &  &  &  & J23 & 22:19 & 2 $\pm$ 1 & -\tablenotemark{\scriptsize l} & Y & N & A\\
			&  &  &  &  & J24 & 22:23 & 3 $\pm$ 1 & 90 $\pm$ 10 & Y & N & A\\
			&  &  &  &  & J25 & 22:33 & 3 $\pm$ 1 & 90 $\pm$ 10 & Y & N & A\\
			&  &  &  &  & J26 & 22:38 & 2 $\pm$ 1 & 100 $\pm$ 10 & Y & N & A\\ 
			\noalign{\smallskip}\tableline \tableline \noalign{\smallskip} 
			Average &  &  &  &  &  &  & 7$\pm$4 & 120 $\pm$ 50 &  \\
			\enddata
			
			\singlespace
			\tablecomments{
				\\\textsuperscript{a}Approximate time at which BMR’s emergence begins in HMI magnetograms. 
				\\\textsuperscript{b}Approximate time at which BMR has finished emerging in HMI magnetograms.
				\\\textsuperscript{c}Emerging BMR's initial location on solar disk in x,y (arcsec).
				\\\textsuperscript{d}Number of emerging BMR’s faint jet spires that we measured.
				\\\textsuperscript{e}Approximate start time of the spire in AIA 193 \AA~images.
				\\\textsuperscript{f}Duration of the spire in AIA 193 \AA~images (for definition see Section \ref{sec:meth}).
				\\\textsuperscript{g}Spire's plane-of-sky speed and its uncertainty measured from summed AIA 193 \AA~images.
				\\\textsuperscript{h}Whether there is base brightening during spire onset and growth in AIA 193 \AA~images [Y (yes), N (no), A (ambiguous)].
				\\\textsuperscript{i}Whether base brightening during spire occurs at a PIL for which there is evidence of flux cancelation [Y (yes), N (no), A (ambiguous)].
				\\\textsuperscript{j}Our judgement of whether the observations indicate the spire is directly driven by the BMR's emergence [Y (yes), N (no), A (ambiguous)].
				\\\textsuperscript{k,n,o,p,q}Spires shoot out repeatedly from the same place on the same PIL.
				\\\textsuperscript{l}Spire too faint to determine speed.
				\\\textsuperscript{m}Spire speed measured from unsummed AIA 193 \AA~images.
			}
			
		\end{deluxetable*}
	\end{longrotatetable}
	
	
	     In the above way, the minifilament-eruption picture for coronal jet production explains why, in observed coronal jets, the pre-jet minifilament traces the PIL of the smaller of the anemone’s two lobes, why the JBP sits near an edge of the jet base, and why the JBP is much smaller than the whole base.  In addition, as \cite{Sterling_2015} point out, the spire’s away-from-the-JBP lateral drift seen in obvious coronal jets is expected in the minifilament-eruption picture for jet production.  In contrast, the \cite{moreno13} three-dimensional numerical simulation of jet production directly driven by external reconnection of an emerging magnetic arch shows spire drift toward the JBP.  Thus, observations have established fairly well that obvious coronal jets in quiet regions and coronal holes are rarely, if ever, directly driven by the emergence of a BMR magnetic arch in a region of largely unipolar far-reaching field.  Instead, observations indicate that the driving of most – maybe all – obvious coronal jets in quiet regions and coronal holes is by a minifilament-flux-rope-cored anemone-lobe eruption that is prepared and triggered by flux cancelation, not flux emergence.  Even so, as is discussed next, it seems physically plausible that as the magnetic arch of a BMR emerges in a region of unipolar flux, emergence-driven reconnection of the arch with ambient far-reaching field might produce at least some jets, or some jet-like events.
	     
	          It is observed that any new BMR is evidently made by the emergence of a magnetic-flux-rope $\Omega$ loop from below the photosphere.  In the birth and growth of a BMR, the centroids of the two photospheric opposite-polarity flux domains begin closest together, continually move apart as the two domains increase in flux and area – growing the connecting magnetic arch in the chromosphere and low corona longer, wider, and higher – and stop moving apart when the flux stops increasing.  That progression is the progression expected for the emergence of a flux-rope $\Omega$ loop \citep{zwaan87,van-Driel-Gesztelyi15,moore20}.
	          
	     When a BMR emerges in a coronal hole, it emerges into the coronal hole’s unipolar open magnetic field.  The open field stems from the majority-polarity flux in the coronal hole.  One foot of the BMR’s emerging magnetic arch is new majority-polarity flux and the other foot is new minority-polarity flux.  The minority-polarity leg of the emerging arch immediately pushes against oppositely directed majority-polarity open field.  That forms a current sheet between them.  So, the minority-polarity leg of the emerging arch soon starts to reconnect with the open field at the current-sheet interface.  The reconnection does two things simultaneously.  It turns the upper part of the reconnected outside of the arch into open field rooted in the outer edge of the BMR’s majority-polarity flux domain.  Simultaneously, the reconnection turns the lower part of the reconnected outside of the arch into a new closed loop connecting the outer edge of the BMR’s minority-polarity flux domain to the ambient extant majority-polarity network flux patch that the reconnected open field was rooted in before its reconnection.  As the magnetic arch continues to emerge, its minority-polarity leg continues to push against and reconnect with oppositely directed majority-polarity open field.  By its emergence-driven external reconnection, the emerging magnetic arch converts itself into a two-lobe magnetic anemone.  One lobe is the still-not reconnected part of the emerged arch.  The other lobe is the magnetic arch/arcade of new closed loops made by the external reconnection of the emerging arch with the ambient open field.  As in the three example emerging BMRs shown in this paper, coronal images of emerging BMRs start showing the field’s expected anemone form within a few hours after the start of emergence is detected in magnetograms having arcsecond resolution and 10 G sensitivity.
	     
	          The emergence-driven external reconnection of the magnetic arch of an emerging BMR, reconnection that makes the anemone’s new lobe – connecting emerged minority-polarity flux to ambient majority-polarity flux – is topologically identical to the emergence-driven external reconnection of the magnetic arch of an emerging BMR in the flux-emergence idea for driving the close-field/open-field reconnection that makes the spire in coronal jets.  The observed growing new lobe of the anemone as a BMR emerges is compelling evidence that emergence-driven external reconnection of the emerging arch does occur.  Therefore, it should be expected that emergence-driven external reconnection of  an emerging BMR’s magnetic arch produces jet-spire-like plasma outflow on the newly reconnected open field that accompanies the production of new loops added to the new lobe.  Even so, flux-emergence-driven jets have not been observed, or, at most, have been much less commonly observed than minifilament-eruption-driven jets.  Why?  A plausible answer is that the external reconnection driven by flux emergence is usually so much weaker than the external reconnection driven by minifilament eruptions that the “spires” resulting from emergence-driven external reconnection are usually so weak that they are not readily noticeable or not easily detectable by present coronal imaging instruments \citep[e.g.][]{Sterling_2015,panesar20}.
	          
	               The purpose of the reported investigation was to search for jet spires made by emergence-driven external reconnection of the magnetic arch in emerging BMRs.  The expected evidence for the production of a spire by emergence-driven external reconnection of an emerging BMR’s arch is sudden brightening and/or sudden growth of the anemone’s new lobe as the spire shoots out.
	               
	          Using full-disk coronal EUV images from the Atmospheric Imaging Assembly (AIA) on \textit{Solar Dynamics Observatory} (\sdo; \citealt{Lemen_2012}) together with full-disk line-of-sight magnetograms from the Helioseismic and Magnetic Imager (HMI) on \sdo, we surveyed coronal holes on the central disk for emerging BMRs that had continuous coverage from the beginning to beyond the end of emergence.  We purposely looked in coronal holes because the dark ambient corona in a coronal hole, especially in AIA 193 \AA\ and 211 \AA\ images, would render the faint spires of weak jets (perhaps ones driven by flux emergence) more visible than if they were to occur in the fog of the ambient corona in non-coronal-hole regions.  We chose emerging BMRs in coronal holes on the central disk so that the arrangement and evolution of the magnetic flux in and around the BMR could be clearly followed in HMI magnetograms having no degradation from near-limb projection effects.  We collected eight such emerging BMRs for close examination, one of which (E7 in Table \ref{tb: params}) is the emerging BMR observed and discussed by \cite{adams14} and \cite{Sterling_2016}. [\cite{adams14} reported that this emerging BMR that occurred in a coronal hole produced no jet visible in AIA coronal EUV images. From closer inspection, \cite{Sterling_2016} reported that this emerging BMR produced a ``weak jet or outflow'' that was discernible in AIA 193 \AA\ images. That ``faint jet'' outflow is much fainter and hence much less noticeable than the coronal EUV jet that occurred elsewhere in the same coronal hole and is the focus of \cite{adams14}. Our knowledge from \cite{adams14} and \cite{Sterling_2016} that emerging BMR E7 produced no coronal EUV jet of normal visibility but did produce a ``faint jet'' spawned the investigation reported here.] We closely tracked each BMR’s jet production during its emergence.  We studied only jets produced during the emergence of their BMR.
	          
	          None of the eight emerging BMRs produced any obvious coronal jets of normal visibility, but each produced one or more jets that we call “faint jets” because the spire is much less visible (much fainter) than in typical coronal jets in AIA 193 \AA\ and 211 \AA\ images.  We studied a total of 26 faint jets from the eight emerging BMRs (Table \ref{tb: params}).  They are so faint that we likely would not have noticed them had we searched only full-disk AIA coronal EUV movies alone (as was done by \citealt{Panesar_2016b}) without zooming in on each BMR.
	          
	          Sixteen of the 26 faint jets evidently originated from where ambient flux was seen to be canceling with opposite-polarity flux of the emerging BMR, and had JBP-like compact brightening on the flux-cancelation PIL.  That is evidence that each of these jets was driven by a flux-rope eruption as in the \cite{Sterling_2015} scenario, with the field in and wrapped around the flux rope having been built and triggered to erupt by flux cancelation, as observed by   \cite{Panesar_2016b,Panesar_2017,Panesar_2018} and by \cite{McGlasson_2019}  for coronal jets driven by minifilament eruptions in quiet regions and coronal holes.  In each of the other 10 faint jets, the spire shoots out as the new lobe of the BMR’s anemone undergoes sudden brightening and/or sudden growth, consistent with production of the spire by emergence-driven external reconnection of the BMR’s magnetic arch.  On the other hand, in each of these jets there is also base brightening at a flux cancelation site or eruptive activity in or near the base that is perhaps the signature of an alternative driver of the production of the faint spire.  For this reason, while each of these 10 faint jets shows evidence of production by the BMR’s emerging arch as in the \cite{yokoyama95} model, each shows evidence of perhaps being produced by a minifilament/flux-rope eruption instead.

	
	\section{Data And Methods}\label{sec:meth}
	
	     To study the cause of faint jet spires from emerging BMRs in on-disk coronal holes, we used EUV images from the 304, 171, 211, 193, and 94 \AA\ channels of \sdo/AIA \citep{Lemen_2012}.  These have 0.6" pixel$^{-1}$, (435 km) and 12 s cadence.  We primarily used 193 \AA\ images because the faint spires were most visible in that channel.  To see and track the photospheric magnetic flux in and around the emerging-BMR faint-jet base, we used line-of-sight magnetograms from \sdo/HMI \citep{scherrer12,schou12}.  These have 0.5” pixel$^{-1}$, 45 s cadence, and a noise level of about 10 G \citep{couvidat16}.

	Using the  JHelioviewer \citep{muller17} service, we identified emerging BMRs in central-disk coronal holes by examining AIA 211 \AA\ images and HMI magnetograms covering most of the disk, viewed at 15-minute time steps.  AIA  211 \AA\ images are good for seeing central-disk coronal holes and good for spotting a coronal-hole BMR’s bright emission, which starts soon after the BMR starts emerging.  Using JHelioviewer, we zoomed in on each BMR for close tracking in higher-cadence 193 \AA\ movies to catch any jets of any strength made during the BMR’s emergence.  We collected eight emerging BMRs that each, during its emergence, did not produce any typical jet spires of normal brightness, but did produce one or more faint jet spires.  Many other closely tracked BMRs produced no jets of any strength during emergence.  A few other closely tracked emerging BMRs produced an obvious jet spire of normal brightness, but these jets were evidently not driven by the emergence of the BMR’s magnetic arch.  Instead, because each of these normal-visibility jets came from an eruption (often carrying a minifilament) from a PIL at which the emerging BMR’s minority-polarity flux was canceling with ambient majority-polarity flux and on which a JBP brightened, each jet was evidently produced by a flux-rope eruption as in the \cite{Sterling_2015} minifilament-eruption scenario for jet production.
	
	For each of the eight BMRs that produced only faint jet spires during emergence, we downloaded \sdo/AIA and \sdo/HMI data from the JSOC cutout service for a small region centered on the emerging BMR. Because each BMR's AIA and HMI cutouts both span  the same x  and y intervals of the full-disk AIA image or full-disk HMI magnetogram to within the width of a pixel, the AIA cutouts are aligned with the HMI cutouts to, at worst, about $\pm$0.5'' in x  and y.  Using SolarSoft routines \citep{freeland98}, for each BMR, we derotated the cutout images and magnetograms to a particular time, so that solar features in the movies show no westward drift from solar rotation.  For each emerging BMR, we made movies at two-minute cadence for the 304, 171, 211, 193, and 94 \AA\ AIA channels, and at three-minute cadence for the HMI magnetograms.  Each movie starts an hour before the onset of flux emergence and ends well after the emergence ends, i.e., hours after the BMR’s flux passes through maximum and starts decaying.  From these movies, we found that the faint jet spires were most visible in AIA 193 \AA\ images, less visible in AIA 171 and 211 \AA\ images, and hardly visible or not visible in AIA 304 and 94 \AA\ images.  Using JHelioviewer, we checked AIA’s other two channels (131 \AA\ and 335 \AA\ ) that, like the 94 \AA\ hot channel, have peak sensitivity to emission from the corona’s hotter plasma at several million Kelvin.  Like the 94 \AA\ channel, they too faintly showed only the brightest of the faint spires.
	
	We then made shorter-time-span movies of the faint jets in AIA 193 \AA\ images at 12 s cadence and in HMI magnetograms at 45 s cadence.  To make the magnetic flux near the noise level more visible in the magnetogram movies, each frame is the sum of two consecutive magnetograms.  To increase the visibility of the faint spires in the 193 \AA\ movies, we used dark-for-bright reversed-color images, and we usually made each frame the sum of two consecutive images.  From these full-cadence 193 \AA\ movies, we obtained the start time, duration, and speed of each of 26 faint jet spires.  These results are listed in Table \ref{tb: params}.
	
	In Table \ref{tb: params}, each faint spire’s speed is the plane-of-sky speed of the spire’s lengthening.  We measured the speed from a time-distance plot of the spire’s intensity along a straight-line cut along the path of the spire in reversed-color AIA 193 \AA\ images; usually we used summed reverse-color AIA 193 \AA\ images (two exceptions are noted in Table \ref{tb: params}).  The measured speed is the slope of a straight line drawn along the track of the spire’s front in the time-distance plot. (For two examples of measurement of the speed of  a coronal jet spire by this method, see Figure 3 of  \cite{Panesar_2016a}.) We repeated the process three times to estimate the uncertainty in the tabulated measured speed, which is the average of the three measured values from the three repetitions.  The spire duration and its uncertainty were obtained visually by stepping frame-by-frame through the duration of the spire in the 193 \AA\ movie.  The duration is the time when the spire starts to fade from maximum visibility minus the time when the spire was first discernible.  The uncertainty in the duration was estimated by choosing the two times from each of three separate frame-by-frame viewings of the movie. [The listed duration is the mean of the three measured durations, and the uncertainty in the listed duration is the uncertainty of the mean, which is the standard deviation of the three measured durations from their mean, divided by the square root of 3.] From the HMI magnetogram movie, we also list in Table \ref{tb: params} the approximate time when each BMR started emerging and the approximate time when the BMR’s emergence ended.
	
	We also made a time-flux plot (such as Figure \ref{fig:flux2}a) from each emerging BMR to show the growth history of flux in the BMR graphically and quantitatively.  We usually measured the emerging BMR’s minority-polarity flux, because it was usually easier to isolate from the ambient majority-polarity flux in the coronal hole than was the BMR’s majority-polarity flux.  To make the time-flux plot, we choose a box within which to integrate all of the BMR’s emerged minority-polarity flux in each magnetogram, a box for which the magnetogram movie shows that no emerged minority-polarity flux migrates out of during the BMR’s emergence.  The plot shows the progression of the BMR’s amount of emerged minority-polarity flux, from none to maximum, as the BMR emerged.

	\section{Results}\label{sec:res}
	\subsection{Overview}\label{sec:over}

     The bare result for each of our 26 faint jets is in the last five columns of Table \ref{tb: params}.  Those  columns are: (1) spire duration, (2) spire speed, (3) whether there was any brightening in the base as the spire shot out, (4) whether there was concentrated brightening at a flux cancelation site at an edge of the base as the spire shot out, and (5) our judgment of whether the spire was made by emergence-driven external reconnection of the BMR’s emerging arch.  In Sections 3.1, 3.2, and 3.3 we describe, as examples, three of the eight emerging BMRs (E2, E4, and E8) and their nine faint jets listed in Table \ref{tb: params}.  In Section 3.4, we briefly describe the other 17 of the 26 faint jets in Table \ref{tb: params}, and discuss the results in Table \ref{tb: params} as a whole.

	
	\begin{figure*}[ht!]
		\includegraphics[width=\textwidth]{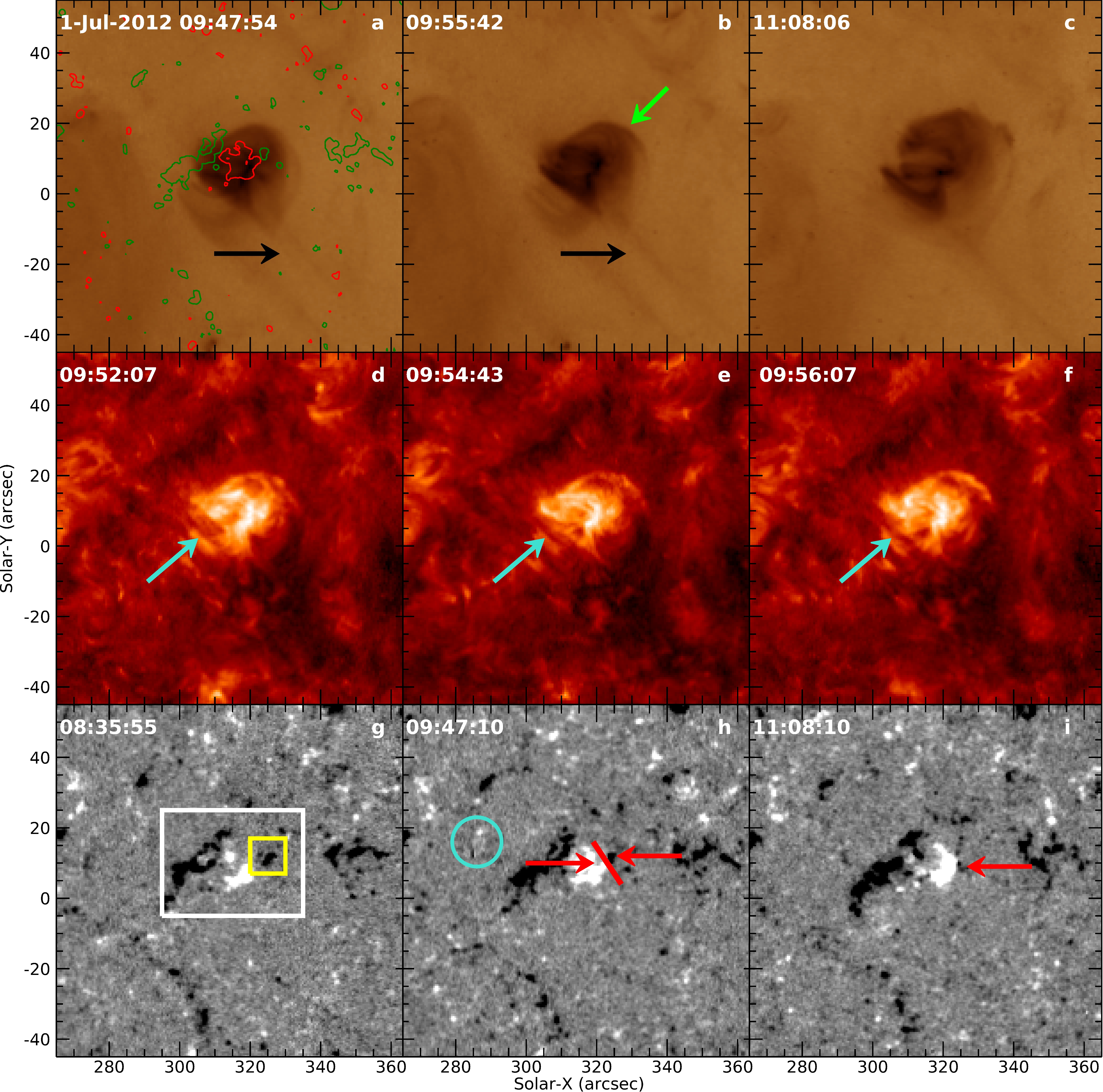}
		\caption{HMI magnetograms of emerging BMR E2 during its burst of emergence in which E2’s three measured faint jet spires occurred, and AIA EUV images of E2 during and after E2’s second faint jet J3.  Panels (a-c) are summed AIA 193 \AA\ reverse-color images.  In panel (a) the black arrow points to J3’s growing faint spire, and HMI magnetogram $\pm$ 50 G contours are overlaid (red = positive, green = negative).  In panel (b) the black arrow points to J3’s faint spire during the spire’s peak visibility, and the green arrow points to the concurrent brightening of the northwest edge of the jet-base anemone's growing western lobe.  Panel (c) is at the end of E2’s third faint jet J4.  Panels (d-f) are AIA 304 \AA\ summed images; the blue arrows point to the surge accompanying J3’s faint spire in panels (a) and (b).  Panels (g-i) are summed HMI magnetograms with saturation level $\pm$75 G.  Panel (g) shows the emerging BMR before the flux cancelation tracked in Figure \ref{fig:flux2}b started.  White and yellow boxes in (g) are the areas measured for the flux evolution plots in Figures 2a and 2b.  Panel (h) is during the flux cancelation; red arrows point to the two opposite-polarity canceling flux patches and the red line marks the PIL between them; the blue circle surrounds the weak mixed-polarity flux at the base of the surge in panels (d-f).  Panel (i) shows E2 after most of the negative flux patch has canceled; the red arrow points to the canceling negative flux patch.  An animation of this figure, from 08:30 UT to 11:30 UT, is available (E2MOVIE); the first panel is summed 193 \AA\ reversed-color images at 12 s cadence, the second panel is summed 304 \AA\ images at 12 s cadence, and the third panel is summed HMI magnetograms (saturation level $\pm$ 75 G) at 45 s cadence.}\label{fig:e2}
	\end{figure*}
	
	\subsection{Emerging BMR E2}\label{sec:canc}

	     Emerging BMR E2 emerged in a coronal hole in which the open field was rooted in flux of negative polarity, the polarity of the majority of the flux in the coronal hole.  During emergence, BMR E2 displays a faint enhanced-brightening narrow extension that is directed toward the southwest, and along which a series of what look like weak outflows occurs. We regard these as spires of a series of weak jets, produced by magnetic changes in the photosphere of the BMR. We identified and measured the spires of three of these weak jets: J2, J3, and J4 of Table \ref{tb: params}. There are several other similar-level brightening enhancements that extend from this emerging BMR and might be faint jet spires, but we analyze only these three here. In Figure \ref{fig:e2}, the top three images are summed reverse-color AIA 193 \AA\ images of E2.  The first two show the faint spire of J3 at 09:47:54 UT as the spire became discernible, and at 09:55:42 UT as the spire was approaching maximum visibility.  The third 193 \AA\ image, at 11:08:06 UT, shows E2 late in the decay of the faint spire of J4.  The middle three images in Figure \ref{fig:e2} are summed AIA 304 \AA\ images during the growth of J3’s faint spire in the reverse-color AIA 193 \AA\ movie (see E2MOVIE (animation of Figure \ref{fig:e2})).  The blue arrows point to a surge-like feature that emanates from near the east side of E2 in the 304 \AA\ movie as J3’s faint spire extends outward in the 193 \AA\ movie.  The surge shoots to the southwest along the direction of the faint spire in the 193 \AA\ images.  The coincidence and co-alignment of the surge with the J3 faint spire suggests that the surge ejection might be the driver of the spire.
	     
 Figures \ref{fig:e2}(g-i) are summed HMI magnetograms of E2 and the ambient predominantly negative-polarity flux in the coronal hole.  They show E2’s emerging flux at 08:35:55 UT, about an hour before faint jet J2, at 09:47:10 UT, about a minute before the start of J3, and at 11:08:10 UT, as the spire of faint jet J4 was fading away.  The blue circle in Figure \ref{fig:e2}h encircles weak mixed-polarity flux at the inferred base of the surge.

\begin{figure}
		\includegraphics[width=\columnwidth]{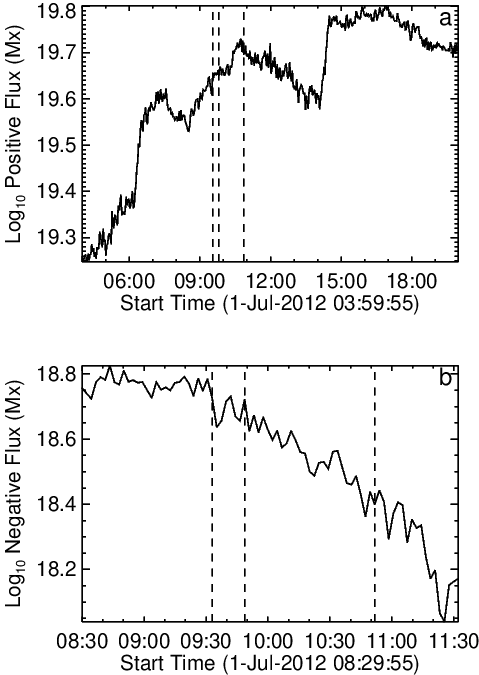}
		\caption{Time-flux plots from emerging BMR E2.  Each plot has log$_{10}$ of the magnetic flux (in Mx) on the y axis and time (UT) on the x axis.  The start times of the three measured faint jet spires from this emerging BMR (J2, J3, J4) are marked by the vertical dashed lines.  Plot (a) is the run of the measured positive (minority-polarity) flux over 16 hours from the start to three hours after the end of E2’s emergence; the plotted flux is the integrated positive flux in the white box in panel (g) of Figure \ref{fig:e2}.  Plot (b) is the integrated negative flux in the yellow box in Figure \ref{fig:e2}g versus time, over the duration of the cancelation of the canceling negative flux patch in panels (g-i) of Figure \ref{fig:e2}.}\label{fig:flux2}  
	\end{figure}

	Figure \ref{fig:flux2}a is the time plot of the total minority-polarity (positive) flux in the white box  in Figure \ref{fig:e2}g, from the beginning to the end of the three-minute cadence 16-hour magnetogram movie for E2.  As the E2 BMR emerged and its two opposite-polarity flux domains spread apart, its minority-polarity positive flux presumably immediately started canceling with any majority-polarity negative ambient flux that it hit, some of which could have been below the 10 G noise level of HMI.  So, the time plot of positive flux in Figure \ref{fig:flux2}a is the time plot of the net positive flux in the white box, the progression of the difference between the amount of new positive flux that had emerged and the amount of new positive flux that had canceled with ambient negative flux.
	
	Figure \ref{fig:flux2}a shows that the emergence of E2 underwent three bursts of emergence as it progressed to peak flux.  The first burst of emergence started about 04:00 UT and ended about 07:30 UT.  For about the next hour, from about 07:30 UT to about 08:30 UT, the net positive flux decreased, indicating that during that interval cancelation of positive flux was out pacing emergence of positive flux.  The second burst of emergence started about 8:30 UT and ended around 10:30 UT.  The third and final burst of emergence started about 14:00 UT and ended around 17:00 UT.  The three vertical dashed lines in Figure \ref{fig:flux2}a mark the start times of the spires of faint jets J2, J3, and J4.  They show that J2 and J3 occurred during the second burst of emergence and J4 started minutes after the end of that rise in net positive flux.  During J4, E2’s magnetic arch probably continued emerging, only not fast enough for the rate of increase of its positive flux by emergence to exceed the rate of decrease of its positive flux by cancelation with invaded ambient negative flux.  The evidence in Figure \ref{fig:flux2}a that J2, J3, and J4 each occurred either during or right after an increase in E2’s net positive flux suggests that each of these faint spines might have been made by a burst of external reconnection of the E2’s emerging magnetic arch with encountered oppositely-directed open field.
	
	The HMI magnetograms in Figure \ref{fig:e2}(g-i) and the magnetogram movie in E2MOVIE show that E2’s emerging magnetic arch had its westward-spreading positive-polarity leading foot directly west of its negative-polarity trailing foot.  That says that the horizontal direction of the arch’s field was eastward.  As can be seen in Figure \ref{fig:e2}, and in movie (E2MOVIE), as the arch emerged it became a magnetic anemone.  That presumably resulted from external reconnection of the minority-polarity (positive) leading leg of the expanding arch with encountered ambient majority-polarity (negative) open field of the coronal hole.  If it were on the upper outside of the leading leg, the external reconnection would build below it the anemone’s western lobe.  The western lobe connects the front side of the emerging BMR’s new positive flux to nearby extant negative flux of the coronal hole.  (The above-mentioned cancelation of the emerging BMR's westward-spreading positive flux with invaded extant negative flux occurs at the PIL inside the western lobe.)  Figure \ref{fig:e2} and E2MOVIE show that as E2’s emerging magnetic arch grows progressively larger in height and east-west extent, evidently, resulting external reconnection of the growing arch builds the western-lobe magnetic arcade progressively larger in height and east-west extent.

	The 193 \AA\ movie in E2MOVIE shows that, starting at about 09:48 UT and ending around 10:00 UT, as J3’s faint spire turns on and grows to maximum visibility, the northwest edge of the western lobe definitely brightens and jumps laterally outward.   Figure \ref{fig:e2}b shows the northwest edge’s brightening and lateral jump in progress. The green arrow in Figure \ref{fig:e2}b points to the new bright edge.  In this way, the 193 \AA\ images show clear evidence that there was a burst of western-lobe-building external reconnection of E2’s emerging magnetic arch that coincided with the production of J3’s faint spire.  The coincidence is the expected signature for J3’s faint spire and the western-lobe’s jump in brightness and extent to have been co-produced by a burst of external reconnection of E2’s emerging magnetic arch, a la \cite{yokoyama95}.  On the other hand, the surge pointed out in the 304 \AA\ images in Figure \ref{fig:e2} also occurred simultaneously with J3’s faint spire.  That coincidence allows the possibility that this surge was responsible both for the surge seen in the 304 \AA\ images and the co-aligned faint spire seen in 193 \AA\ images, and that the burst of external reconnection that made the western lobe’s burst of growth was not strong enough to make a discernible faint spire.  Therefore, we judge that the burst of western-lobe-building external reconnection of E2’s emerging arch perhaps made J3’s faint spire, but the coincidence of the 304 \AA\ surge renders J3’s spire-production mechanism ambiguous.
	
	    In Figure \ref{fig:e2}g, the yellow box isolates a clump of negative-polarity coronal-hole flux that, at about an hour before the first of E2’s three faint jets, is being approached from the east by E2’s westward-spreading emerging positive flux.  At 09:47 UT  (Figure \ref{fig:e2}a,h)  the western edge of the new positive flux had reached the negative flux clump and was canceling it.  Figure \ref{fig:e2}i shows that by the end of E2’s third faint jet the encroaching new positive flux had canceled most of the negative flux clump.
	     
	Figure \ref{fig:flux2}b is a time plot of the negative flux in the yellow box in Figure \ref{fig:e2}g, from about an hour before E2’s first faint jet, J2, to about half an hour after E2’s third faint jet, J4.  The three vertical dashed lines in Figures 2a and 2b mark the start times of the faint spires of J2, J3, and J4.  On these plots, they show that each of E2’s three faint jets happened when the negative flux clump was canceling with the west edge of E2’s emerging positive flux.  That raises the question of whether any of these three faint jets were driven by an eruption of a flux rope that was built and triggered to erupt by flux cancelation at the PIL between the negative flux clump and E2’s invading positive flux.  The occurrence of a JBP-like compact brightening on the PIL at the time of a faint spire would suggest that such a flux-rope eruption did drive the production of the spire by \cite{Sterling_2015}.  Absence of coincident JBP brightening on the PIL would be evidence that the faint spire probably was driven some other way.
	
	In the 193 \AA\ movie in E2MOVIE, from about 09:33 UT to around 09:40 UT, as J2’s faint spire starts and grows to maximum visibility, the west edge of E2’s western lobe brightens and jumps westward. (In the frame at 09:39:06 in that movie, a green arrow points to that lobe's brightened west edge.)  That is clear evidence that a burst of western-lobe-building external reconnection of E2’s emerging arch coincided with the production of J2’s faint spire.  That coincidence is the expected sign for production of the faint spire a la \cite{yokoyama95}, by the burst of emerging-arch external reconnection that made the burst of western-lobe growth.  On the other hand, the 193 \AA\ movie shows a JBP-like compact bright feature on the flux-cancelation PIL, starting about 09:30 UT and ending around 09:35 UT.  It has peak brightness at about 09:32 UT, about a minute before J2’s faint spire becomes discernible. (In the frame at 09:33:42 in that movie, a green arrow points to that compact bright feature, and a black arrow points to the brightening faint spire.) Compact brightening on the flux-cancelation PIL at the onset of the spire is a sign that the faint spire might have been produced a la \cite{Sterling_2015}, by the eruption of a flux rope that was build and triggered to erupt by the flux cancelation \citep{Panesar_2016b}.  In addition, the 304 \AA\ movie shows that a surge similar to, from the same place, and in the same direction as the surge that accompanied the faint spire of J3, accompanied the faint spire of J2. (In the frame at 09:32:19 in that movie, a blue arrow points to the surge that is occurring during J2.)  This is a sign that, like J3’s faint spire, J2’s faint spire might have resulted from the coincident surge ejection from the mixed-polarity site circled in Figure \ref{fig:e2}h.  Thus, while there is evidence that J2’s faint spire was made a la \cite{yokoyama95}, there is simultaneous evidence that J2’s spire was made in either of the above two alternative ways.  We therefore judge that the evidence for how J2’s spire was produced is ambiguous.
	
	In the 193 \AA\ movie, there is no JBP-like compact brightening on the flux-cancelation PIL after the one for J2, none at the times of J3 and J4.  The absence of any compact on-PIL brightening in the E2’s western lobe at the times of J3 and J4 is an indication that neither of these faint spires was the result of an eruption of a flux rope that was built and triggered to erupt by flux cancelation at the PIL of the western lobe.
	
	In the 193 \AA\ movie, from about 10:52 UT to about 11:03 UT, as J4’s faint spire starts and grows to maximum visibility, the northwest edge of E2’s western lobe brightens and jumps laterally outward to the northwest. (In the frame at 10:53:42 in that movie, a green arrow points to that lobe's brightening west edge, and a  black arrow points to the J4 faint spire.) That is, like the western-lobe growth during J2 and J3, during J4’s faint spire, the western lobe has a burst of growth that is presumably from a burst of western-lobe-building external reconnection of E2’s emerging magnetic arch.  So, as do J2 and J3, J4 shows evidence indicating that its spire was made a la \cite{yokoyama95}, by a burst of external reconnection of E2’s emerging arch.  But, again as for J2 and J3, the 304 \AA\ movie shows that a third surge, from the place circled in Figure \ref{fig:e2} and in the same direction of the other two surges, shoot out in synchrony with, and along the path of J4’s spire. (In the frame at 10:56:07 in that movie, a blue arrow points to the surge that is occurring during J4.) So, perhaps J4’s spire was produced by the coincident surge ejection instead of a la \cite{yokoyama95}.  For that reason, we judge the evidence for how the J4 faint spire was produced to be ambiguous.
	
	The  193 \AA\ movie of E2MOVIE shows that the three brightest faint jet spires produced by E2 during  the three-hour span of  the movie were the spires of faint jets J2, J3, and J4. Each of these three faint jets was selected for close inspection and measurement because of the good visibility of its spire.   Upon closer inspection we fount that  each of these spires happened during  the  corresponding surge that is pointed out in the 304 \AA\ movie. In our judgment, for each of the three faint jets J2, J3, and J4, the coincidence of  the spire  with the corresponding surge together with the alignment  of  the surge  with the spire is sufficient evidence to make  the case of the spire ambiguous, even though  the coincident  jump in brightness and outer extent of the jet-base anemone's external lobe is consistent with the spire having been made by a burst of emergence-driven external reconnection of E2's emerging magnetic arch. During  the time interval from 10:23  UT to 10:36 UT, the 304 \AA\ movie shows a similar surge from  the same place and having the same duration as the three surges  that accompanied J2, J3, and J4, but in the 193 \AA\ movie there  is no corresponding faint spire of brightness similar to that of  the faint spires of J2, J3, and J4. This  raises the possibility  that  each surge accompanying J2, J3, and J4 was merely coincidental  rather than causal. Even so, the coincidence and alignment of each of the accompanying surges  with the spires of J2, J3, and J4, in our opinion, makes the cause of  each of these three faint jet  spires ambiguous.
			\subsection{Emerging BMR E4}\label{sec:canc1}
			
\begin{figure*}[htb!]
				\includegraphics[width=\textwidth]{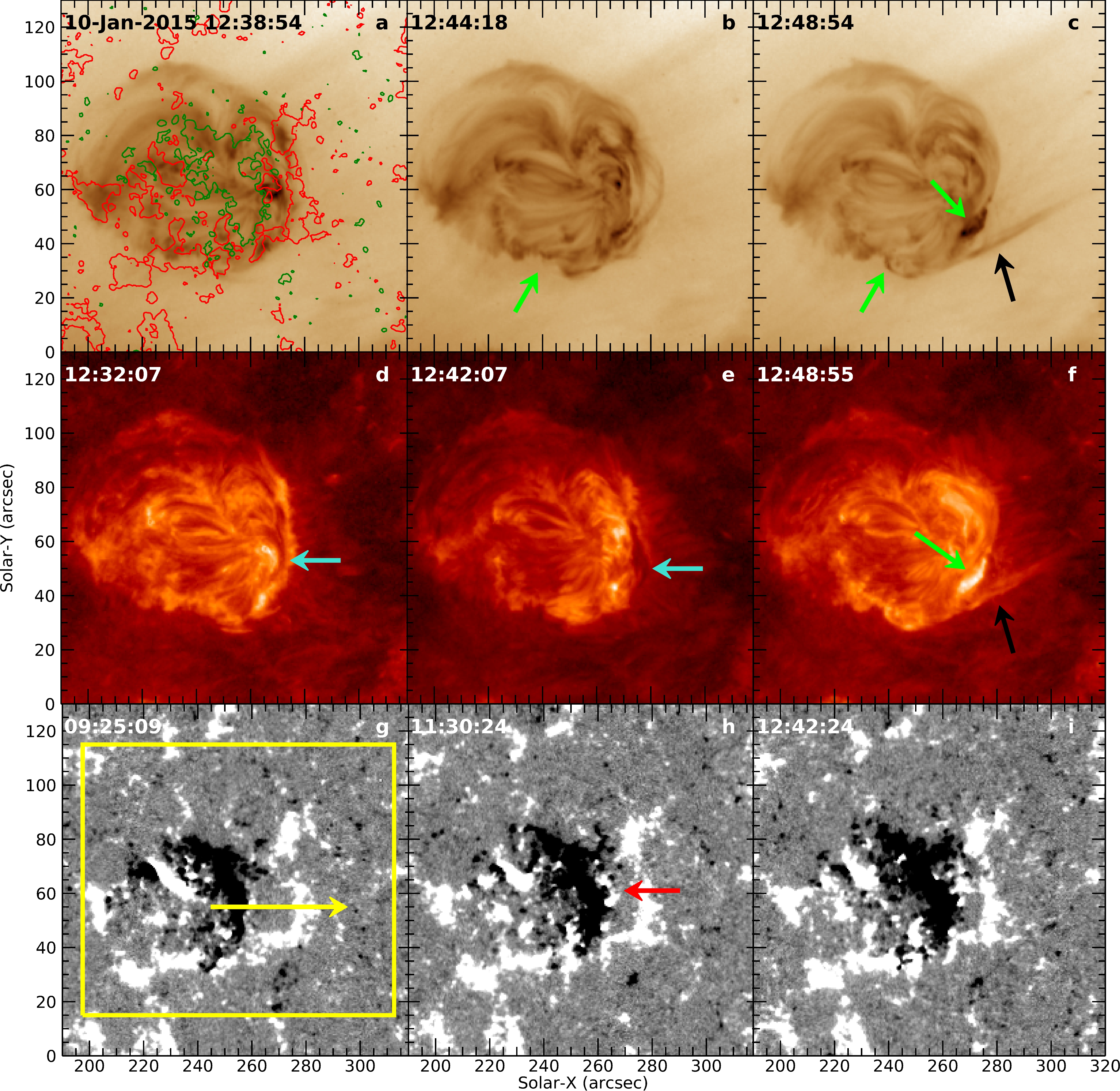}
				\caption{HMI magnetograms of emerging BMR E4 sampling the emergence of E4 over a three and a quarter hour interval ending minutes before the onset of E4’s faint jet J7, and AIA EUV images of E4 minutes before and during jet J7.  Panels (a-c) are summed AIA 193 \AA\ images.  Panels (d-f) are summed AIA 304 \AA\ images.  Panels (g-i) are summed HMI magnetograms with saturation level $\pm$75 G.  In panel (a), the overlaid HMI magnetogram contours are for $\pm$ 50 G (red = positive, green = negative).  Panel (b) is during the onset of the J7 faint jet. The green arrow in (b) and the lower green arrow in (c) point to bright feet of hot loops added to the outside of E4's emerging magnetic arch by the minifilament-eruption-driven external reconnection of the anemone's western lobe. The blue arrow in panel (d) points to the minifilament about 10 minutes before it erupted.  The blue arrow in panel (e) points to the erupting minifilament.  The black arrows in panels (c) and (f) point to J7’s spire at peak visibility.  The upper green arrow in (c) and the green arrow in (f) point to the JBP on the PIL from which the minifilament erupted.  The yellow box in panel (g) is the area in which the negative flux was measured for the time-flux plots in Figure \ref{fig:flux4}a and Figure \ref{fig:flux4}b.  The yellow arrow in panel (g) shows the spatial range and spatial direction of the HMI time-distance map in Figure \ref{fig:flux4}c.  At roughly an hour before the minifilament erupts, the red arrow in panel (h) points to a positive flux clump that is canceling at the minifilament’s PIL.  The magnetogram in panel (i) is during the minifilament’s eruption, and shows that some of the positive flux clump at the PIL obviously cancelled in the hour interval between panel (h) and panel (i).  An animation of this figure, from 09:20 UT to 13:00 UT, is available (E4MOVIE); cadences of the AIA EUV images and HMI magnetograms in E4MOVIE are the same as for E2MOVIE.
				}\label{fig:e4}
			\end{figure*}

			     Emerging BMR E4 emerged in a coronal hole in which the majority of the magnetic flux had positive polarity.  Of our eight emerging BMRs, E4 emerged the longest (about 2 days), grew to be the biggest (attained the size of a large supergranule, spanning $\sim$ 40,000 km), and emerged the most flux ($\sim$10$^{21}$ Mx), enough to make small sunspots.  As the 193 \AA\ (T $\sim$ 1.5 $\times$ 10$^{6}$ K) coronal images, 304 \AA\ (T $\sim$ 5 $\times$ 10$^{4}$ K) cooler-transition-region images, and photospheric magnetograms in Figure \ref{fig:e4} show, E4’s emerging magnetic arch was aligned roughly east-west.  The horizontal direction of the field in the emerging arch was westward, giving the flux in the emerging arch’s leading foot negative polarity.
			     
			The two-minute-cadence 193 \AA\ movie and the three-minute-cadence magnetogram movie both cover the entire 2-day emergence of E4.  About five hours after E4 begins emerging in the magnetogram movie, the 193 \AA\ movie begins to show that some of the negative-polarity leading leg of E4’s emerging arch has reconnected with ambient open field to make a small lobe of loops connecting some of E4’s westward-spreading negative flux to weak positive flux in front (west) of it.  The weak (near noise level) positive flux is in a channel between the westward-advancing front of E4’s negative flux and a north-south network lane of concentrated positive flux farther west.  Together, the westward-connecting small lobe and E4’s emerging arch form a lopsided two-lobed magnetic anemone surrounding E4’s negative flux.  As E4’s arch continues to emerge, the westward-connecting lobe grows larger, presumably from further external reconnection of the emerging arch, but remains much smaller in east-west span than the eastward-connecting emerging-arch lobe.  As the arch emerges and its negative-polarity front moves westward, the growing westward-connecting lobe envelops progressively more of the narrowing channel of weak positive flux.  That is, the western feet of the westward-connecting lobe march toward the north-south network lane.
			
			E4MOVIE (animation of Figure \ref{fig:e4}) is an animation from which we selected the E4 images and magnetograms in Figure \ref{fig:e4}.  E4MOVIE shows the evolution of E4 at full cadence in 193 \AA\ images, 304 \AA\ images, and magnetograms for about three and a half hours on 2015 January 10, from about 09:30 UT to about 13:00 UT.  At the beginning of the movies in E4MOVIE, the magnetograms show that E4’s negative flux is roughly in the shape of a north-south crescent having its two horns bent back to the east, as can be seen in  Figure \ref{fig:e4}g.  The southwest front of the westward-advancing negative-flux crescent is in contact with the eastern end of a roughly east-west lane of positive network flux.  At the beginning of the movies, the front of the north-south middle of the crescent is still well separated from the roughly north-south positive network lane in front (west) of it.  By this time, the western feet of the growing westward-connecting lobe have almost reached the eastern edge of the north-south network lane, and so that lobe envelops practically all of the narrowing channel of weak flux.  Inside the westward-connecting lobe seen in the 193 \AA\ images, a wide swath of dark striations seen in the 304 \AA\ images conforms to and covers the wide channel of weak positive flux between the front of the negative-flux crescent and the north-south positive-flux network lane.  We take this 304 \AA\ dark swath to be a wide minifilament or minifilament channel along the PIL of the westward-connecting lobe.  At the end of the magnetogram movie in E4MOVIE, as can be seen in the third magnetogram in Figure \ref{fig:e4}, the weak-flux channel has narrowed to less than half as wide as it was at the start of the movie.  Correspondingly, the  304 \AA\ image (at 12:32:07 UT) in Figure \ref{fig:e4}d shows, near the end of the movies in E4MOVIE, the minifilament or minifilament channel tracing the PIL and the weak-flux channel along the front of the negative-flux crescent had become much narrower than it was at the beginning of the movies.  By this time, as the magnetogram contours in Figure \ref{fig:e4}a (at 12:38:54 UT) show, the western feet of the westward-connecting lobe have marched into the north-south network lane.
			
	At 09:30 UT, the magnetogram movie shows (Figure \ref{fig:e4}g), at the north edge of the shaft of the yellow arrow, there is a small patch of positive flux at the edge of the west front of the negative-flux crescent.  That positive flux started emerging there about three hours earlier.  Positive flux continues to emerge there and near there during the rest of the three and a half hours of the magnetogram movie in E4MOVIE.  In Figure \ref{fig:e4}h, the red arrow points to the on-PIL compact emerging positive flux.

	\begin{figure}
		\includegraphics[width=\columnwidth]{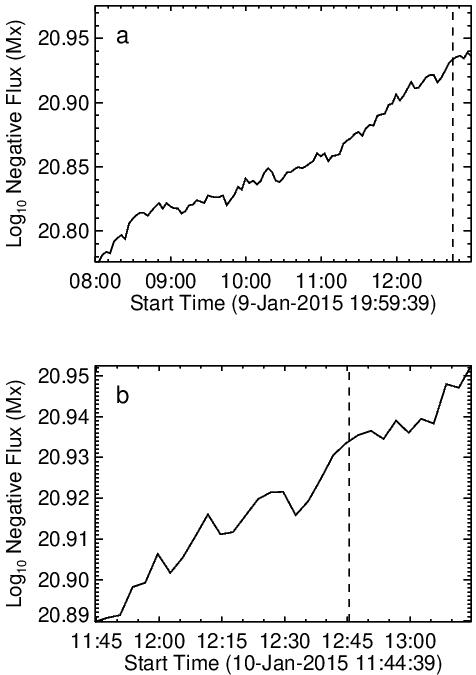}\vspace{0.3cm}
		\includegraphics[width=\columnwidth]{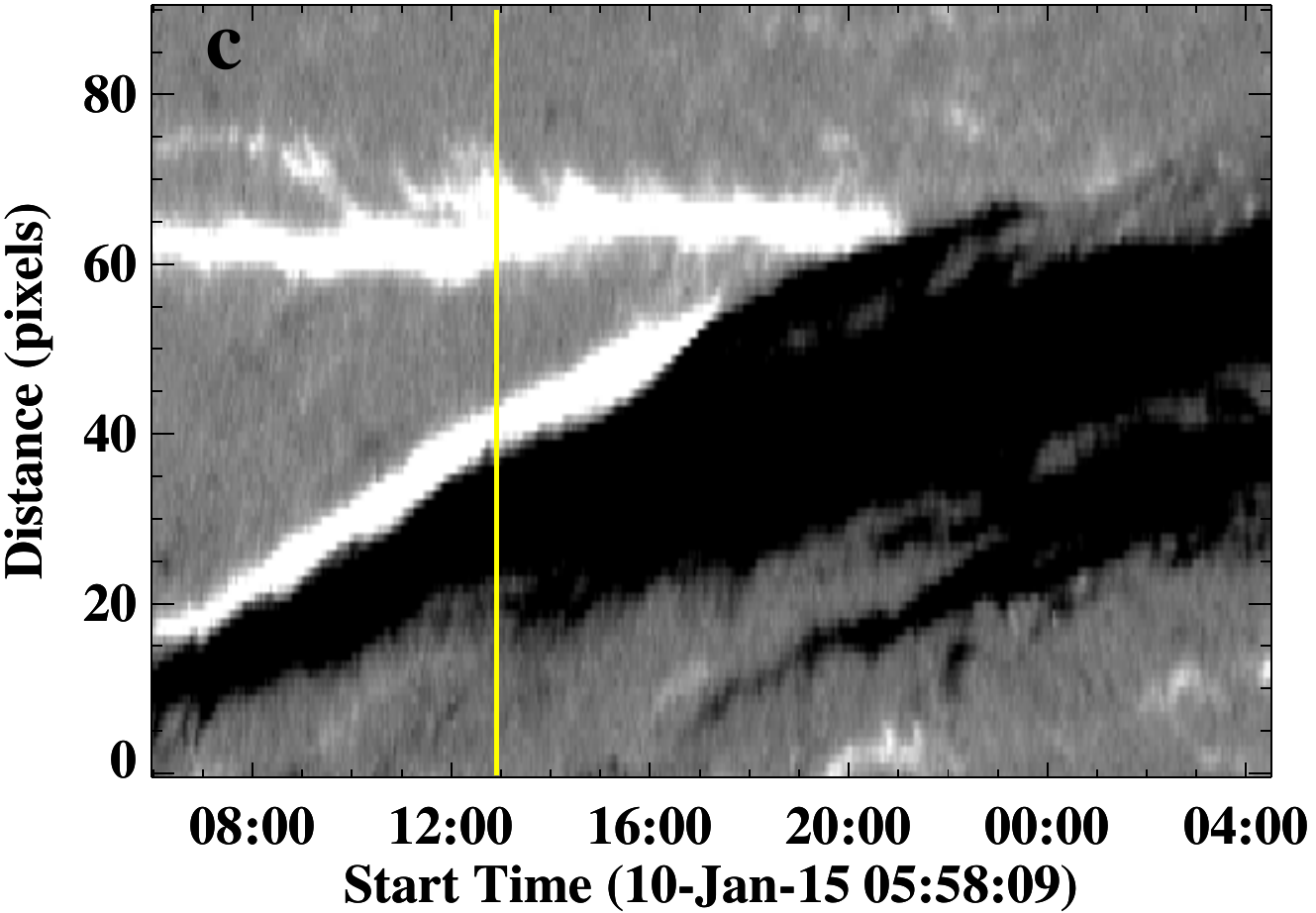}
		\caption{Two time-flux plots and an east-west-cut time-distance plot from HMI magnetograms of emerging BMR E4.  Plot (a) is the integrated negative flux in the yellow box in Figure \ref{fig:e4}g over 5 hours, ending 15 minutes after J7.  Plot (b) is the integrated negative flux in the same box, but is a closer look covering only 1.5 hr, starting 1 hr before before J7.  Panel (c) is the magnetogram time-distance map from along the yellow arrow in Figure \ref{fig:e4}g.  The dashed vertical line in panels (a) and (b) and the yellow vertical line in panel (c) mark the start time of J7’s faint spire.}\label{fig:flux4} 
	\end{figure}
	
	The wide minifilament/minifilament channel present at 09:30 UT inside the western lobe of the E4 anemone shows that the field in that channel had strong left-handed shear by then.  The striations in the 304 \AA\ minifilament/minifilament channel show that the field threading and directing the striations is fairly flat (roughly horizontal) and runs roughly north-south, filling the weak-flux channel in front (west) of the PIL along the front edge of the negative-flux crescent.  From the sweep of the striations, the field in the minifilament/minifilament channel evidently connects positive flux along the southern end of the negative-flux crescent across the PIL to negative flux near the PIL farther north along the crescent (as in typical solar filaments/filament channels \cite[e.g.][]{panesar20b}).  Perhaps much of the left-handed magnetic shear was generated by E4’s negative flux reconnecting to positive network flux on its south side early in E4’s emergence, and then shearing past that positive network flux as E4’s emergence continued to drive E4’s negative flux westward.
	
	After 09:30 UT, as E4’s negative flux continued to advance westward, flux-cancelation reconnection of any sheared field rooted in positive flux in the weak-flux channel on the west side of the PIL plausibly could concentrate the shear along the PIL and build a minifilament-holding flux rope a la \cite{balle89}, and perhaps eventually trigger that flux rope to erupt and drive the production of a jet.  In any case, as E4MOVIE and Figure \ref{fig:e4} show, by 12:30 UT, the wide minifilament/minifilament channel had greatly narrowed.  By 12:42 UT, the minifilament is seen to be erupting in the 304 \AA\ images (blues arrows in Figure \ref{fig:e4}d,e).  The minifilament eruption produced J7's faint spire, which is pointed to by the black arrows in Figure \ref{fig:e4}c,f.
	
	During the rest of E4’s emergence, after J7 ejected\label{key}, as E4’s negative flux continued to move westward and cancel encountered ambient positive flux, a few more minifilament-eruption-type eruptions that erupted from along the flux-cancelation PIL each produced a faint jet spire.  Each of these eruptions is obviously similar to the minifilament eruption that produced E4’s faint jet J7 shown in Figure \ref{fig:e4}.  Because of their similarity, we selected only J7 for closer examination of the production of the faint spire by the minifilament eruption.  We present the production of J7 here in detail as a good example of faint-jet production that definitely was not driven by external reconnection of the BMR’s emerging magnetic arch, because it was obviously driven by a minifilament eruption a la \cite{Sterling_2015}.
	
 Figure \ref{fig:flux4}a is a time plot of E4’s net negative flux in the yellow box  in Figure \ref{fig:e4}g.  It starts at 08:00 UT, about 12 hours after the emergence of E4 became discernible in the magnetograms, and ends 5 hours later, at 13:00 UT.  Figure \ref{fig:flux4}b gives a closer look at E4’s net-negative-flux time profile during the hour and a half from 11:45 UT to 13:15 UT.  The vertical dashed line in panels (a) and (b) of Figure \ref{fig:flux4} marks the start time of J7’s faint spire.  These two plots show that J7 ejected after nearly 17 hours of E4’s emergence, right at the end of a 15-minute burst in the growth of E4’s net negative flux.  That timing might be taken to be an indication that J7’s faint spire was made by a burst of emergence-driven external reconnection of E4’s emerging magnetic arch.
	
Figure \ref{fig:flux4}c is a time-distance plot of the magnetic flux along the length of the yellow arrow  in Figure \ref{fig:e4}g.  The tail end of the arrow is at the bottom of the plot and the tip is at the top.  The yellow vertical line in the plot marks the start time of J7’s faint spire, at 12:45 UT.  In the magnetograms, the yellow-arrow cut, from east to west, runs through E4’s negative-flux crescent, the compact positive flux patch at the PIL on the front of the crescent, and the north-south lane of positive flux.  The time-distance plot displays the westward advance of the front of E4’s negative flux toward the nearly stationary positive-polarity network lane.  The plot shows that, along the yellow-arrow cut, the negative-flux front finally reached and canceled the network lane about 7 hours after J7 occurred.  The plot also shows that, on the yellow-arrow cut, the positive flux patch at the PIL on the front of the negative-flux crescent gradually grew wider for about 4 hours after 06:00 UT, and then kept roughly constant width until finally being canceled by the negative-flux crescent during the two hours from about 16:00 UT to about 18:00 UT.
	
	From the time-distance plot, along with the narrowing of the 304 \AA\ minifilament/minifilament channel between the advancing negative-flux front and the positive-flux network lane, we conjecture that the minifilament flux rope was built and triggered to erupt in two steps as follows.  First, flux-cancelation reconnection low above the PIL between legs of the minifilament-channel sheared field rooted in positive flux in the weak-flux channel and opposite legs of that sheared field rooted in the advancing negative flux near the PIL built a minifilament flux rope that sat above and to the west of the PIL.  Then, some of the flux rope’s field-line tethers rooted near the PIL under the minifilament were cut by flux-cancelation tether cutting driven by the compact flux emergence at the PIL, as envisioned by \cite{moore92}.  We conjecture that this tether cutting triggered the eruption of the flux rope and its enveloping lobe of E4’s anemone.
	
	In any case, the minifilament eruption in Figure \ref{fig:e4} and in E4MOVIE shows clear signatures of external and internal reconnection of the E4-anemone lobe that envelops the erupting minifilament flux rope, clear signatures of the external and internal reconnection expected in the \cite{Sterling_2015} minifilament-eruption idea for coronal jet production.  The eruption-driven external reconnection of the erupting lobe is expected to produce new hot loops on the outside of the E4’s emerging magnetic arch.  The eastern bright feet of these expected new hot loops  are seen along the southeast edge of the emerging-arch lobe of E4’s anemone in the 193 \AA\ images of Figure \ref{fig:e4}b,c.  At the same time, as the minifilament flux rope erupts and the faint spire appears, the expected JBP brightens on the flux-cancelation PIL under the erupting minifilament.  The green arrows  in Figure \ref{fig:e4}c,f point to the JBP.  Therefore, we judge that the evidence is unambiguously conclusive that J7’s faint spire was not produced by external reconnection of E4’s emerging magnetic arch, but was produced by external reconnection of the other lobe of E4’s anemone, external reconnection driven by the eruption of the minifilament flux rope in the core of that lobe.

	\subsection{Emerging BMR E8} \label{sec:emer}
	
\begin{figure*}[htb!]
		\includegraphics[width=\textwidth]{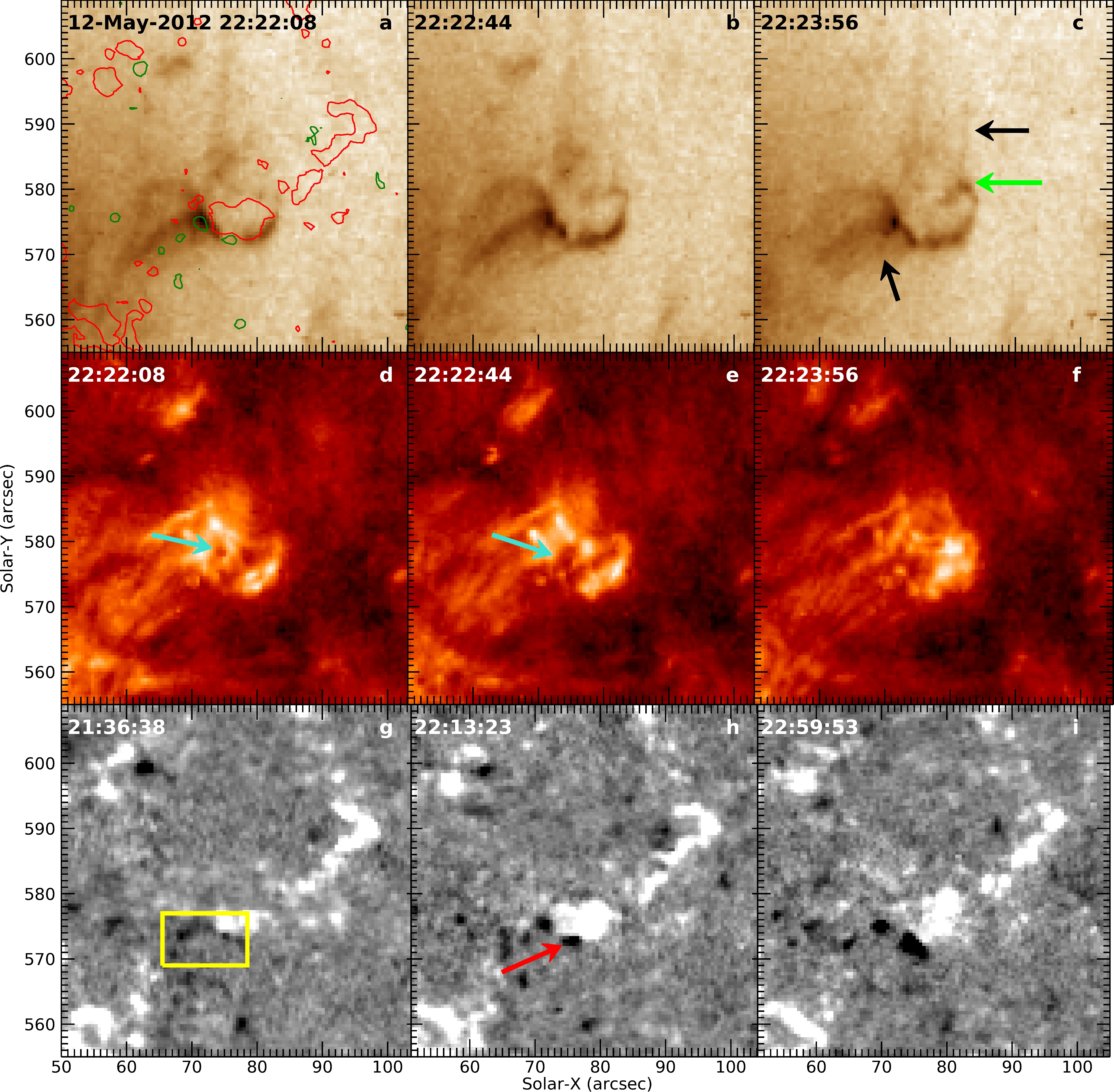}
		\caption{HMI magnetograms of emerging BMR E8 sampling the emergence of E8 at onset of emergence and at the beginning and end of E8’s period of fastest emergence, and AIA EUV images of E8 in the early onset and growth of E8’s third faint jet J24.  Panels (a-c) are summed AIA 193 \AA\ reversed-color images.  Panels (d-f) are summed AIA 304 cancelation images.  Panels (g-i) are summed HMI magnetograms with saturation level $\pm$ 75 G.  Panels (a) and (d) show E8 roughly one minute before the onset of J7’s faint spire.  The overlaid magnetogram contours in panel (a) are for $\pm$50 G (red = positive, green = negative).  In panel (b), J7’s faint spire and bright base are starting to become faintly visible.  In panel (c), the upper black arrow points to the growing faint spire,  the green arrow points to the brightening at the spire’s foot, and the lower black arrow points to a newly brightened loop of E8's external lobe.  The blue arrow in panels (d) and (e) points to what appears to be an erupting minifilament.  Panel (g) shows the very onset of E8’s flux emergence, when E8’s negative flux was first becoming discernible in the summed HMI magnetograms.  The yellow box in panel (g) is the area in which the negative flux was measured for the time-flux plots in Figure \ref{fig:flux8}.  The red arrow in panel (h) points to an emerging negative flux patch.  Panel (i) shows E8’s emerged flux at the end of E8’s period of rapid emergence.  An animation of this figure, from 11:40 UT to 23:00 UT, is available (E8MOVIE); the format and cadence are the same as in E2MOVIE and E4MOVIE.
		}\label{fig:e8}
	\end{figure*}

	Emerging BMR E8 emerged in the Sun’s northern hemisphere, in a coronal hole in which the majority of the magnetic flux had positive polarity.  In the magnetogram movie in E8MOVIE (animation of Figure \ref{fig:e8}), E8 starts emerging at about 21:40 UT on 2012 May 12 and reaches its maximum emerged flux in only four hours.  Because the emerging positive-polarity flux is northwest of the emerging negative-polarity flux, the horizontal direction of the field in the emerging magnetic arch is southeast.  So, E8’s magnetic arch was tilted by roughly 45 degree from east-west, and the polarity of its leading flux domain was opposite that which larger BMRs in the northern hemisphere usually had in 2012 (early in the maximum phase of Solar Cycle 24).  The magnetogram in Figure \ref{fig:e8}g shows E8 as it was becoming discernible in the magnetograms. The magnetogram in Figure \ref{fig:e8}g is at 21:36:38 UT, virtually simultaneous with the 21:40:23 UT first frame of the magnetogram movie in E8MOVIE. Figure \ref{fig:e8}h shows E8 early in E8’s phase of most rapid emergence, at a time (22:13:23 UT) early in faint jet J22, the first of E8’s five faint jets that we measured.  Figure \ref{fig:e8}i, at 22:59:53 UT, shows E8 at the end of its most rapid phase of emergence, about 20 minutes after faint jet J26, the last of E8’s five faint jets that we measured. As did emerging BMR E2, E8 made several other faint jet spires, but we selected only these five to measure. We selected these five faint spires for close inspection and  measurement of  their duration  and speed because we judged these five  faint  spires to be somewhat more distinctly  visible than the others in E8’s 193 Å movie.  Even so, two of these five (J22 and J23) were  each too faint for its speed to be measured from its time-distance plot.
	
	
\begin{figure}
	\includegraphics[width=\columnwidth]{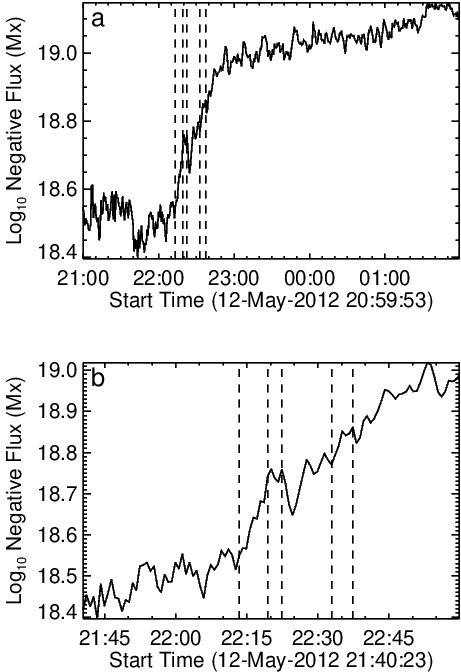} 
	\caption{Time-flux plots of emerging BMR E8.  Each plot has log$_{10}$ of the negative flux (in Mx) on the y axis and time (UT) on the x axis.  Plot (a) is the integrated negative flux in the yellow box in Figure \ref{fig:e8}g over 5 hours, starting 40 minutes before the onset of emergence and ending 10 minutes after maximum negative flux in the yellow box.  Plot (b) gives a closer look at the rise of the integrated negative flux in the yellow box during E8’s period of fastest emergence, which started at about 22:05 UT and ended about 50 minutes later.  In each plot, the five vertical dashed lines mark the start times of the spires of E8’s five faint jets (J22, J23, J24, J25, J26).}\label{fig:flux8}
\end{figure}

	Figure \ref{fig:flux8} shows two time plots of the growth of E8’s minority-polarity (negative) flux in the yellow box  in Figure \ref{fig:e8}g.  The plot in Figure \ref{fig:flux8}a starts about 40 minutes before E8 started emerging, and ends five hours later, about 20 minutes after E8’s net negative flux in the yellow box reached its maximum.  The plot in Figure \ref{fig:flux8}b starts at 21:40 UT, which is when the measured negative flux in the yellow box started increasing, and ends at 23:00 UT, which is roughly at the end of E8’s phase of most rapid emergence.  The start times of the faint spires of E8’s five faint jets that we measured (J22, J23, J24, J25, J26) are marked by the five vertical dashed lines in each plot in Figure \ref{fig:flux8}.  They show that each of these faint jets occurred during E8’s phase of most rapid emergence, and hence are prospective candidates for being produced by emergence-driven external reconnection of E8’s emerging magnetic arch.
	
	In the 193 \AA\ movie in E8MOVIE, the first coronal signs of E8’s magnetic field are not coronal loops of the emerging magnetic arch, but some faint coronal loops of the E8 magnetic anemone’s other lobe.  That lobe is E8’s external lobe.  The external lobe is presumably built by external reconnection of the emerging arch with encountered oppositely-directed open field of the coronal hole.  It connects E8’s minority-polarity (negative) flux to extant clumps of majority-polarity (positive) flux southeast of the emerging arch.  Before reconnection with the emerging arch, a strand of coronal-hole open field is rooted in the positive flux that will be in the positive end of the external lobe’s new loop that will be made by that reconnection.  After reconnection, the reconnected strand of open field is rooted in the northwest end of E8’s emerged positive flux.  The negative-polarity end of the external lobe’s new loop is rooted in the southeast end of E8’s emerged negative flux.  From the start of E8MOVIE at 21:40 UT until about 22:10 UT, in the 193 \AA\ movie, the emerging arch is hardly seen, if at all, while a few loops of the external lobe are intermittently faintly visible.  During this time in the 304 \AA\ movie, the emerging arch is a bright blob that is sometimes laced with blurry dark striations aligned southeast-northwest.  The striations are reminiscent of the arch filament system that would be observed in H$\alpha$ images of an emerging magnetic arch aligned southeast-northwest \citep[e.g.][]{bruzek77}.  The 304 \AA\ striations and the absence of noticeable emerging-arch loops in the 193 \AA\ images suggest that during this time, as the emerging-arch magnetic loops welled up from chromospheric to coronal heights, they were quickly reconnected to make new loops of the external lobe before coronal heating could make them visible in the 193 \AA\ images.

	In the 193 \AA\ movie, the spire of faint jet J22 becomes discernible at about 22:13 UT, reaches its maximum height and brightness at about 22:15 UT, and has become fainter by 22:17 UT.  The spire extends north from its slightly brighter base near the north edge of E8’s positive flux.  During the occurrence of the spire from 22:13 UT to 22:17 UT, new bright loops appear in the external lobe extending southeast from E8’s negative flux. (In the frame at 22:14:43 in the 193\AA\ movie in E8MOVIE, the two black arrows point to J22's spire, the upper green arrow points to J22's bright base, and the lower green arrow points to a new bright loop of E8's external lobe.) The spire’s occurrence during the rapid emergence of E8’s magnetic arch, the spire’s base setting, the base brightening, and the spire’s occurrence at the same time as new bright loops in the external lobe are all consistent with J22’s spire having been produced a la \cite{yokoyama95} by a burst of emergence-driven external reconnection of the emerging arch.  On the other hand, during J22, the magnetograms in E8MOVIE show weak negative flux along the edge of E8’s positive flux, near the base of the spire. In the frame at 22:14:53 UT in the magnetogram movie in E8MOVIE, a blue circle encircles that weak negative flux. The enhanced brightness of the spire’s base and the presence of mixed-polarity flux at the base allows the possibility that the spire was not driven a la \cite{yokoyama95}, but was driven a la \cite{Sterling_2015} by the eruption of a tiny flux rope that was built and triggered to erupt by flux cancelation at the base of the spire.  Therefore, while there is good evidence suggesting that J22’s faint spire was produced by emergence-driven external reconnection of E8’s emerging arch, because there is mixed-polarity flux and discernible brightening at the base of the spire, we judge that the evidence for whether J22 was produced a la \cite{yokoyama95} is ambiguous.
	
	In the 193 \AA\ movie, like J22, E8’s other four faint jets that we measured during E8’s rapid emergence (J23, J24, J25, J26 in Table \ref{tb: params}) extend north from their bases along the north edge of E8’s positive flux, as expected if each were produced a la \cite{yokoyama95}.  J23’s base and J25’s base are at about the same place as J22’s base, J24’s base is farther west, and J26’s base is between the J22-J23-J25 base and J24’s base.  As during J22, during each of the other four faint jet spires, one or more new loops brighten in E8’s external lobe to the southeast, as expected if the spire were produced a la \cite{yokoyama95}.  On the other hand, as for J22, during each of the other four faint jets, the magnetograms in E8MOVIE show small patches of weak negative flux near the base of the spire, at the outside edge of E8's positive flux, and the 193 \AA\ images show base brightening. (In the frame at 22:20:19, 22:23:55, 22:34:19, and 22:38:19, respectively for J23, J24, J25, and J26 in the 193\AA\ movie, the black arrows point to the spire, the upper green arrow points to the spire's bright base, and the lower green arrow points to a brightening or newly brightened loop of E8's external lobe.) Therefore, on the same basis as for J22, we judge that the evidence for whether spires of J23, J24, J25, and J26 were produced a la \cite{yokoyama95} is ambiguous.
	
	The 193 \AA\ images and 304 \AA\ images in Figure \ref{fig:e8} are snapshots of E8 just before and during the onset of faint jet J24, which starts at about 22:23 UT.  The first 193 \AA\ image and first 304 \AA\ image (both at 22:22:08 UT) show E8 about a minute before the start of J24.  The blue arrow in the first 304 \AA\ image points to a “minifilament” dark feature inside E8’s emerging magnetic arch.  Thirty-six seconds later, the second 193 \AA\ image shows J24’s base faintly brightening and J24’s faint spire starting to extend from its base.  At the same time, the second 304 \AA\ image shows that the “minifilament” (pointed to by the blue arrow) is erupting.  In the 304 \AA\ movie, it appears that the “minifilament eruption” is a confined eruption that is arrested within E8’s emerging magnetic arch. (We take this dark feature to be a minifilament and take its observed  eruptive motion in E8's 304 \AA\ movie to be a confined eruption of magnetic field carrying the dark-feature plasma because the dark feature's form and motion are similar to some of the minifilament eruptions reported in \cite{Sterling_2015}, whereby minifilament eruptions that produced a non-blowout narrow spire were largely confined within  the jets's closed-field base.) In the last 193 \AA\ image in Figure \ref{fig:e8}, a minute after the second 193 \AA\ image, J24’s base (pointed to by the green arrow) is growing bigger and brighter and the faint spire (pointed to by the upper black arrow) is growing longer.  In the second and third 193 \AA\ images in Figure \ref{fig:e8}, a new loop (pointed by the lower black arrow in Figure \ref{fig:e8}c) is brightening on the southwest side of the arch of extant bright loops of E8’s external lobe.  In the 193 \AA\ movie, the new loop grows more distinct as J24’s faint spire grows to maximum visibility at about 22:25 UT, two minutes after the third 193 \AA\ image in Figure \ref{fig:e8}.  The occurrence of the new bright loop simultaneously with J24’s faint spire is the expected signature for the production of J24’s spire by a burst of external reconnection of E8’s emerging magnetic arch.
	
	The occurrence of the “minifilament eruption” at the onset of J24 raises the possibility that J24’s spire was driven by a burst of external reconnection of the emerging arch but the reconnection was not driven by the arch’s emergence.  The coincidence suggests that the reconnection was instead driven by the “confined-minifilament-eruption” convulsion of the magnetic arch [(\cite{Sterling_2015} report that such convulsions can be the cause of so-called ``standard jets''(albeit the spire-producing confined minifilament eruptions reported by \cite{Sterling_2015} were not in the  emerging lobe of  the jet-base anemone but in the anemone's other lobe, the external lobe made by external reconnection of the  emerging lobe within ambient open field)]. The “minifilament” eruption at the onset of J24 is further evidence making the cause of J24’s spire ambiguous.  
	
	If J24's faint spire was made by a burst of external reconnection of E8's emerging magnetic arch and that reconnection was not driven by the emergence of that arch's magnetic field but by the minifilament-eruption internal convulsion of already-emerged field in that arch, then J24's spire was made neither a la \cite{yokoyama95} nor a la \cite{Sterling_2015}, but a la a third way that has been envisioned and/or simulated by  \citep[e.g.,][]{Moore_2010,archontis13,pariat15,liu16}. In this third way, the magnetic field in the jet-base anemone's emerging lobe, by being already twisted before it emerges or by getting sheared and twisted by photospheric flows during and/or after it merges, holds free magnetic energy, some of which can be released in a magnetic convulsion (like that in a minifilament eruption) that makes a jet spire by driving a burst of external reconnection of the emerging lobe.  That is, in this concept, the emerging magnetic lobe drives a burst of spire-making external reconnection not by its emergence a la \cite{yokoyama95}, but by an internal convulsion of twisted/sheared field inside it. Of our 26 faint jets, E8's J24 is the only one in which we noticed prospective evidence (such as the minifilament eruption in J24) that the spire was possibly driven in this third way rather than either a la  \cite{yokoyama95} or a la \cite{Sterling_2015}. 
	
	In each of E8's five jets (J22-J26), instead of the jet's base being - as we have supposed - a tiny bipole that  is formed, made explosive, and triggered to erupt to drive the spire and  brighten the base of the spire  by  convergence and partial cancelation of the tiny negative flux patch with E8's positive-polarity flux, the  tiny bipole  of the base of the spire is conceivably an emerging bipole in which there is a minifilament flux rope that was not  built by  flux cancelation and that erupts to drive the spire and the spire's base brightening. The negative flux at the  feet  of these jets is  too near the noise level of the magnetograms for us to discern whether this alternative was the  case. However, that this third way is a possibility for making each of these jets is an additional reason for us to judge that whether  each  was made a la \cite{yokoyama95} or some other way  is ambiguous, i.e., to judge that each should have an A in the last column of Table \ref{tb: params}.	

	\subsection{Overview of the Results in Table \ref{tb: params}} \label{sec:results}
	
	     So far, of the eight emerging BMRs in Table \ref{tb: params}, we have presented in detail the observations of three of them (E2, E4, E8) and the observations of their nine faint jets for which we measured the spire durations and speeds listed in Table \ref{tb: params} (E2’s J2, J3, J4; E4’s J7; E8’s J22, J23, J24, J25, J26).  E4 is an example of an emerging BMR whose measured faint jet spire was clearly driven by a minifilament-flux-rope eruption prepared and triggered by flux cancelation.  E2 and E8 are examples of emerging BMRs whose measured faint jet spires showed evidence of possibly being made a la \cite{yokoyama95} via emergence-driven external reconnection of the emerging magnetic arch, but also showed evidence of possibly being made a la  \cite{Sterling_2015} and \cite{Panesar_2016b} via reconnection driven by a minifilament-flux-rope eruption prepared and triggered by flux cancelation.
	     
	Each of the 10 measured faint jets (J1, J2, J3, J4, J21, J22, J23, J24, J25, J26) produced by four of the emerging BMRs in Table \ref{tb: params} (E1, E2, E7, E8) showed evidence of possibly being produced a la \cite{yokoyama95}.  Each of these 10 jets also showed evidence for possibly being produced a la \cite{Sterling_2015}, \cite{Panesar_2016b,Panesar_2018}, and \cite{McGlasson_2019}, as in the presented observations for the eight faint jets produced by E2 and E8.  Therefore, for each of these 10 faint jets, the entry in the last column of Table \ref{tb: params} is A, for the evidence being ambiguous for whether the spire was produced a la \cite{yokoyama95}.  Each of the other 16 measured faint jets in Table \ref{tb: params}, the 16 faint jets produced by the other four emerging BMRs in Table \ref{tb: params} (E3, E4, E5, E6), showed no evidence of possibly being produced a la \cite{yokoyama95}, but did show evidence for possibly being produced by a minifilament-flux-rope eruption.  Each of these 16 faint jet spires originated from a PIL for which there is evidence of flux cancelation and at which there is concentrated brightening during the spire.  The evidence for ongoing flux cancelation at the PIL is a steep gradient from negative to positive flux across the PIL and/or opposite-polarity flux converging on the PIL, as in Figure \ref{fig:flux4}c for E4.  Therefore, for each of these 16 faint jets, the entry in the last column of Table \ref{tb: params} is N, for the evidence being unambiguous that the spire was not produced a la \cite{yokoyama95}.
	
	The durations of the 26 faint jet spires in Table \ref{tb: params} range from about 2 minutes to about 15 minutes.  The average duration is 7 minutes.  These durations of the spires of faint jets seated in coronal holes and observed in AIA 193 \AA\ images are similar to the spire durations found in previous studies of coronal jets of normal visibility seated in coronal holes, both in coronal EUV images \citep{Panesar_2018} and in coronal X-ray images \citep{Savcehva_2007}.   Our coronal-hole faint-jet-spire durations are also similar to the durations found for the spires of EUV coronal jets of normal visibility observed not in coronal holes but in quiet regions \citep{Panesar_2016b}.
	
For our 10 faint  spires for each of  which the evidence is ambiguous for whether the spire was produced a  la  \cite{yokoyama95}, the average duration and its 1-$\sigma$ uncertainty are 6.5$\pm$1.6 min. For our 16 faint spires for each of which the evidence  is conclusive that  the spire was not produced a la \cite{yokoyama95},  the average duration  is only marginally  longer, 7.8$\pm$0.7 min. That  is, the difference in the two average  duration is not statistically significant. 

	For the 22 faint jet spires for which we were able to measure the speeds listed in Table \ref{tb: params}, the speed ranges from about 35 km s$^{-1}$ to about 235 km s$^{-1}$.  The average speed is 120 km s$^{-1}$.  The standard deviation of the 22 speeds from the average speed is $\pm$50 km s$^{-1}$. Our measured spire speeds of faint EUV jets in coronal holes are similar to the measured spire speeds of normal-visibility X-ray jets and EUV jets in coronal holes \citep[e.g.,][]{Savcehva_2007,Panesar18b}, quiet regions \citep[e.g.,][]{Panesar_2016b}, and active regions \citep[e.g.,][]{Shimojo&H_1996}.
	
	For the 8 faint spires for each of which the speed could be measured and the evidence is ambiguous for whether the spire was produced a la \cite{yokoyama95}, the average speed  and its 1-$\sigma$ uncertainty  are 91$\pm$2.3 km s$^{-1}$.  For the 14 faint spires for each of which the speed could be measured and the evidence is conclusive that the spire was not produced a la \cite{yokoyama95}, the average speed is 134$\pm$13.4 km s$^{-1}$.  The difference  in average speed is definitely statistically significant.  Under the assumption that the 8 ambiguously-produced faint spires were actually produced a la \cite{yokoyama95}, the difference suggests the following for the reconnection that makes faint jet spires.  The faint-spire-producing reconnection of a lobe of the jet-base anemone with open field tends to be driven more slowly when it is driven by emergence of the lobe than when it is driven by core-field eruption of the lobe.  This suggestion is compatible with (1) the observation that spreading of the feet of the emerging magnetic arch of an emerging BMR is usually slower than $\sim$ 1 km s$^{-1}$ \citep{panesar20,moore20}, and (2) the observation that for normal-visibility EUV jets in central-disk coronal holes, the growing spire usually does not start to become discernible in AIA coronal EUV images until the speed of the driving erupting minifilament surpasses a few km s$^{-1}$ \citep{panesar20}. These two observations suggest that minifilament eruptions often drive the spire-making external reconnection fast enough to make an obvious spire of normal brightness, and  that external reconnection directly driven by  the  emergence of a jet base's emerging magnetic arch is seldom, if ever, fast enough to make an  obvious spire of normal brightness. Our observations indicate that emergence-driven  external reconnection might sometimes be fast enough to make a faint spire, but is seldom fast enough to make an  obvious spire of normal brightness, By an  ``obvious spire of normal brightness'', we mean one as obvious  and bright as  the spires of  the coronal jets reported by \cite{Sterling_2015}, \cite{Panesar_2016b,Panesar_2017,Panesar_2018} and \cite{McGlasson_2019}. Of course, for each of the faint jets that showed evidence of the spire possibly being produced directly by the emergence of the BMR's magnetic arch, the observations allow that the arch's external reconnection that made the spire was instead driven by a minifilament-flux-rope eruption from a site of weak flux cancelation. For this alternative, the slower spire speeds could have resulted simply from the driving minifilament-flux-rope eruption in the slower-spire faint jets being weaker than in the faster-spire faint jets. 

	Similarly, it is reasonable that in any of our 26 faint jets that were actually driven by a minifilament-flux-rope eruption, the spire was much fainter than in minifilament-eruption-driven jets of obvious visibility. The relatively smaller and weaker mixed-polarity flux elements at the bases of faint jets would plausibly make minifilament flux ropes that are smaller and weaker than those that erupt to make more obviously visible coronal jets. Eruption of these comparatively small/weak flux ropes would likely produce spires that are less energetic and fainter, slower-moving, and shorter-lived than for jets of normal visibility.

	\section{Summary and Conclusion}
	
	     It is observationally well established that the magnetic field of a BMR emerging in a larger area of predominantly unipolar magnetic field in a quiet region or coronal hole is a two-lobed magnetic anemone.  One lobe is the emerging magnetic arch of the emerging BMR.  The other lobe is the external lobe of closed loops made by external reconnection of the emerging arch with ambient far-reaching field rooted in majority-polarity magnetic flux near the minority-polarity foot of the emerging arch.  The external lobe connects some of the BMR’s emerged minority-polarity flux to nearby extant clumps of majority-polarity flux.
	     
	It is observed that coronal jets often emanate from the magnetic anemones of emerging BMRs in quiet regions and coronal holes.  It is widely accepted that the spires of such jets are produced by external reconnection of a lobe of the anemone with ambient oppositely-directed far-reaching field that the lobe presses against to drive the reconnection.  It is now observationally well established that, in at least many obvious coronal jets from emerging-BMR anemones in quiet regions and coronal holes, the driver of the spire-making external reconnection is an eruption of a minifilament flux rope built and triggered to erupt by flux cancelation in the core of the external lobe of the jet-base BMR anemone \citep{Sterling_2015,Panesar_2016b,Panesar_2017,Panesar_2018} and \cite{McGlasson_2019}.  On the other hand, several numerical simulations of coronal jet production had previously shown that spire-making reconnection could plausibly be made by the emerging magnetic arch’s emergence-driven external reconnection that builds the external lobe of the BMR’s anemone \citep{yokoyama95,nishizuka08, Moreno-Insertis_2008, moreno13}, as originally suggested by \cite{Shibata_1992}.  From observations of the onset of EUV coronal jets in central-disk coronal holes, \cite{panesar20} have proposed that the reason that obvious coronal jet spires produced by emergence-driven reconnection of an emerging BMR’s emerging magnetic arch are seldom observed is that the emergence is seldom fast enough to drive external reconnection strongly enough to make an obvious spire.  This suggests that emergence-driven external reconnection of a BMR’s emerging magnetic arch might produce spires that are so faint that they are seldom noticed in coronal images, but might sometimes be discernible in adequately favorable circumstances.
	
	We searched for faint jet spires made a la \cite{yokoyama95} by examining the production of the spires of 26 coronal EUV jets that emanated from eight emerging-BMR magnetic anemones in central-disk coronal holes.  During emergence of the BMR, none of the eight anemones produced any obvious jets of normal visibility.  Each of the 26 spires was much fainter than the spires of typical previously-studied obvious coronal jets.  Each was so faint that it would not have been noticed in a movie of full-disk AIA coronal EUV images.  Each would not have been found had we not closely followed the jet-base magnetic anemone in a zoomed-in movie of AIA coronal EUV images, and had the spire not been viewed against the dark background of a coronal hole.
	
	Because each of the 26 faint jets occurred on the central disk, HMI line-of-sight magnetograms showed, without confusion from near-limb projection effects, the arrangement and evolution of the magnetic flux in and around the jet-base emerging-BMR anemone.  The magnetograms showed whether there was a PIL at the foot of a faint spire, key evidence for judging which of two main possibilities for producing the faint spire was more likely.  One main possibility for spire production is a la \cite{yokoyama95} by emergence-driven external reconnection of the BMR’s emerging magnetic arch.  The other main possibility for spire production is a la \cite{Sterling_2015} and \cite{Panesar_2016b} by the BMR’s external lobe’s external reconnection driven by the lobe’s eruption driven by a minifilament-flux-rope eruption prepared and triggered by flux cancelation in the lobe’s core. A third possibility  is  that the spire is produced by external reconnection of the jet-base anemone's emerging lobe driven in another way, namely, by an internal convulsion  of that lobe's magnetic field rather than by  the emergence of that lobe. We found tentative evidence for this third possibility in only one of our 26 faint jets, E8's J24.
	
	Along with the HMI magnetograms, we viewed the onset and growth of each faint jet spire in an AIA 193 \AA\ reversed-color 12-second-cadence movie.  The expected signature for the production of the faint spire a la \cite{yokoyama95} is a burst of brightening and/or growth of the emerging BMR’s external lobe during the onset and growth of the spire.  The expected signatures for the production of the faint spire a la \cite{Sterling_2015}, \cite{Panesar_2016b}, and \cite{McGlasson_2019} are: (1) a PIL at the foot of the spire, (2) eruption of a cool-plasma minifilament and/or surge from the PIL during the onset and growth of the spire, (3) EUV brightening concentrated on the PIL during the onset and growth of the spire, and (4) evidence of ongoing flux cancelation at the PIL before, during, and after the spire happens.
	
	Sixteen faint jet spires were all of the measured spires produced by four of our eight emerging BMRs.  Each of these 16 faint spires was not accompanied by the expected signature for spire production a la \cite{yokoyama95}.  Each of these 16 faint spires had a PIL at its foot and was accompanied by one or two or all three of the other three expected signatures for spire production a la \cite{Sterling_2015,Panesar_2016b}, \cite{McGlasson_2019}.  Therefore, for each of these 16 faint spires we judge that the evidence is conclusive that the spire was not produced by emergence-driven reconnection of the BMR’s emerging magnetic arch.  We surmise that the spire was produced by external reconnection of an external lobe of the BMR, and that this spire-producing external reconnection was driven by an eruption of that lobe’s minifilament-flux-rope core field.
	
	The other 10 faint jet spires were all of the measured spires produced by the other four of our eight emerging BMRs.  Each of these 10 faint spires was accompanied by the expected signature for spire production a la \cite{yokoyama95}.  Therefore, each of these 10 faint spires may have been produced by emergence-driven external reconnection of the BMR’s emerging magnetic arch.  To the contrary, each of these 10 faint spires also had a PIL at its foot and was accompanied by one or two of the other three expected signatures for spire production a la \cite{Sterling_2015}, \cite{Panesar_2016b}, and \cite{McGlasson_2019}.  Therefore, the observations for each of these 10 faint spires allow the possibility that (1) the BMR’s emerging arch’s burst of external reconnection that produced the burst  of brightening and/or growth of the BMR’s external lobe did not produce a corresponding spire that was bright enough to be faintly seen in the AIA 193 \AA\ images, and (2) the observed faint spire was instead produced by a burst of external reconnection of an erupting lobe of the BMR’s anemone, driven by a minifilament-flux-rope eruption in the core of that lobe.  Therefore, for each of these 10 faint spires, we judge that the evidence is ambiguous for whether the spire was produced a la \cite{yokoyama95}.
	
	For the 26 measured faint jet spires from eight emerging BMRs, our conclusion is that in only the 10 faint spires from four of the emerging BMRs was the emergence-driven external reconnection of the emerging magnetic arch perhaps vigorous enough to make the observed faint spire. Moreover, we cannot rule out that these faint jets were made instead by weak minifilament eruptions, in the fashion that we know larger more obvious jets are frequently made. Thus our results indicate that the external reconnection driven by an emerging BMR’s emerging arch is seldom, if ever, driven fast enough and hence strongly enough to produce an obvious spire of normal visibility.  This is why nearly all obvious coronal jets are observed to be driven by a minifilament-flux-rope eruption that is prepared and triggered by flux cancelation, and hardly any are observed to be produced by emergence-driven external reconnection of the magnetic arch of an emerging BMR.

	
	\acknowledgments
	This research was conducted thanks to the REU program funded by NSF and hosted by the University of Alabama in Huntsville and NASA’s Marshall Space Flight Center. This material is based on work supported by the National Science Foundation under Grant No. AGS-1460767.  N.K.P  acknowledges support from HGI grant  (80NSSC20K0720) and NASA’s \sdo/AIA (NNG04EA00C). AIA is an instrument onboard the Solar Dynamics Observatory, a mission for NASA’s Living With a Star program. R.L.M and A.C.S acknowledge the support from the NASA HGI program. We are indebted to the \sdo/AIA and \sdo/HMI teams for providing the high resolution data. \sdo\ data are courtesy of the NASA/\sdo\ AIA and HMI science teams. We thank  the referee for comments and insights that  clarified our findings.
	
	\bibliography{faint-jets}{}
	\bibliographystyle{aasjournal}
	
\end{document}